\documentclass[11pt]{article}
\usepackage{xypic}
\usepackage{mathrsfs}
\usepackage{amssymb}
\usepackage{amsfonts}
\usepackage{amsmath}

\def\ben{\begin{enumerate}}
\def\een{\end{enumerate}}

\def\bit{\begin{itemize}}
\def\eit{\end{itemize}}

\def\0{\leqno}

\input xy
\xyoption{all}
\input{tcilatex}
\begin{document}

\begin{center}
{\Large {LAGRANGIAN MECHANICS}\ \ \bigskip }

{\Large {ON GENERALIZED LIE ALGEBROIDS}\bigskip }


\textbf{by }

\textbf{CONSTANTIN M. ARCU\c{S} }
\end{center}

\bigskip





%

\begin{abstract}
A solution for the Weinstein's Problem in the general framework of
generalized Lie algebroids is the target of this paper. We present the \emph{%
mechanical systems} called by use, \emph{mechanical }$\left( \rho ,\eta
\right) $\emph{-systems, Lagrange mechanical }$\left( \rho ,\eta \right) $%
\emph{-systems} or \emph{Finsler mechanical }$\left( \rho ,\eta \right) $%
\emph{-systems} and we develop their geometries. We obtain the canonical $%
\left( \rho ,\eta \right) $\emph{-}semi(spray) associated to a mechanical $%
\left( \rho ,\eta \right) $-system. The Lagrange mechanical $(\rho ,\eta)$%
-systems are the spaces necessary to develop a Lagrangian formalism. We
obtain the $(\rho ,\eta)$-semispray associated to a regular Lagrangian $L$
and external force $F_{e}$ and we derive the equations of Euler-Lagrange
type. A new point of view over classical and modern results about Lagrangian
Mechanics are presented. \ \ \bigskip\newline
\end{abstract}

\tableofcontents

\ \

\section{Introduction}

The notion of Lagrange space was introduced and studied by J. Kern $\left[ 7%
\right] $ and R. Miron $\left[ 12\right] .$ The geometry of Lagrange spaces
was extensively examined by geometers and physicists from Canada, Germany,
Hungary, Italy, Japan, Romania, Russia and USA. Many international
conferances were devoted to debate this subject, proceedings and monographs
were published $\left[ 4,5,13,14,15,18\right] .$ In the classical sense, a
regular Lagrangian on $TM$ is a smooth function $%
\begin{array}[b]{ccc}
TM & ^{\underrightarrow{~L\ }} & \mathbb{R}%
\end{array}%
$ such that the Hessian matrix with entries%
\begin{equation*}
\begin{array}[b]{c}
g_{ij}\left( x,p\right) =\frac{1}{2}\frac{\partial ^{2}H\left( x,y\right) }{%
\partial y^{i}\partial y^{j}}%
\end{array}%
\end{equation*}%
is everywhere nondegenerate on $TM$ (or on a domain of $TM$) and a Lagrange
space is a pair $L^{n}=\left( M,L\right) ,$ where $L$ is a regular
Lagrangian (see $\left[ 15\right] $).

We know that a geodesic of the Lagrange space $L^{n}$ is an extremal curve
of the action integral%
\begin{equation*}
\begin{array}[b]{c}
I\left( c\right) =\dint\limits_{0}^{1}L\left( x\left( t\right) ,\frac{%
dx\left( t\right) }{dt}\right) dt.%
\end{array}%
\end{equation*}

This is, in fact, a solution of the Euler-Lagrange system of equations:%
\begin{equation*}
\begin{array}[b]{cc}
\dot{x}^{i}=\frac{dx^{i}\left( t\right) }{dt}, & \frac{d}{dt}\left( \frac{%
\partial L}{\partial \dot{x}^{i}}\right) -\frac{\partial L}{\partial x^{i}}%
=0,%
\end{array}%
\end{equation*}%
where $\left( x^{i}\left( t\right) ,i\in \overline{1,m}\right) $ are the
local coordinates of the point $c\left( t\right) .$

This system is equivalent to
\begin{equation*}
\begin{array}[b]{c}
\frac{d^{2}x^{i}}{dt^{2}}+2G^{i}\left( x,\frac{dx}{dt}\right) =0.%
\end{array}%
\end{equation*}

Here $G^{i},i\in \overline{1,m}$ are the components of a semispray that
generates a notable nonlinear connection, whose coefficients are given by
\begin{equation*}
\begin{array}[b]{c}
N_{k}^{i}=\frac{\partial G^{i}}{\partial y^{k}}.%
\end{array}%
\end{equation*}

The case when $L$ is square of a function on $TM,$ positively, $1$%
-homogeneous with respect to the velocity $y^{i},$ provides an important
class of Lagrange spaces called Finsler spaces.

We know the \textbf{Weinstein's Problem}: \bigskip

\begin{quotation}
{\small Develop a Lagrangian formalism directly on the given Lie algebroid
similar to Klein's formalism for ordinary Lagrangian Mechanics (see }$%
{\small [8]}${\small ).}

\bigskip
\end{quotation}

This problem was formulated by A. Weinstein in $[23]$, where the author gave
the theory of Lagrangians on Lie algebroids and obtained the Euler-Lagrange
equations using the dual of a Lie algebroid and the Legendre transformation
defined by a regular Lagrangian. In $[10]$, P. Liberman showed that such a
formalism is not possible if one consider the tangent bundle of a Lie
algebroid as space for developing the theory. Using the prolongation of a
Lie algebroid over a smooth map, E. Martinez solved the \textbf{Weinstein's
Problem} in $[11]$ (see also $[6,9]$).

A Lagrangian description of Mechanics on Lie algebroids was extensively
studied by many authors. (see $\left[ 16,17,19,20,21,22\right] $)

The generalized Lie algebroid is a new notion necessary to obtain a new
class of (linear) connections in Ehresmann sense.(see $[1]$) The notions of
\emph{IDS} and \emph{EDS} for Lie algebroids presented in $[3]$ was natural
extended to generalized Lie algebroids in $\left[ 2\right] $. In particular
there are obtained a new point of view over exterior differential calculus
for Lie algebroids.

In this paper we propose to solve the \textbf{Weinstein's Problem} in the
general framework of generalized Lie algebroids.

In the Sections $3,4,5$ and $6$ we set up the basic notions and terminology.
In Section $7$ we introduce the \emph{mechanical systems} called by use,
\emph{mechanical }$\left( \rho ,\eta \right) $\emph{-systems, Lagrange
mechanical }$\left( \rho ,\eta \right) $\emph{-systems} or \emph{Finsler
mechanical }$\left( \rho ,\eta \right) $\emph{-systems.} In Section $8$ we
present the \emph{\ canonical }$\left( \rho ,\eta \right) $\emph{-semispray
associated to mechanical }$\left( \rho ,\eta \right) $\emph{-system }$\left(
\left( E,\pi ,M\right) ,F_{e},\left( \rho ,\eta \right) \Gamma \right) $%
\emph{\ and from locally invertible }$\mathbf{B}^{\mathbf{v}}$\emph{%
-morphism }$\left( g,h\right) .$ If $\left( \rho ,\eta \right) =\left(
Id_{TM},Id_{M}\right) $, $\left( g,h\right) =\left( Id_{E},Id_{M}\right) $,\
and $F_{e}\neq 0$, then we obtain\ the canonical semispray associated to a
connection $\Gamma $ presented by I. Bucataru and R.~Miron in~$[5]$. Also,
we obtain \emph{the canonical }$\left( \rho ,\eta \right) $\emph{-spray
associated to mechanical system }$\left( \left( E,\pi ,M\right)
,F_{e},\left( \rho ,\eta \right) \Gamma \right) $\emph{\ and from locally
invertible }$\mathbf{B}^{\mathbf{v}}$\emph{-morphism }$(g,h).$

The Section $9$ is dedicated to study the geometry of Lagrange mechanical $%
\left( \rho ,\eta \right) $-systems. These mechanical systems are the spaces
necessary to solve the \textbf{Weinstein's Problem }in the general framewok
of generalized Lie algebroids. We determine and we study the $\left( \rho
,\eta \right) $-semispray associated to a regular Lagrangian $L$ and
external force $F_{e}$ which are applied on the total space of a generalized
Lie algebroid and we derive the equations of Euler-Lagrange type. In
particular, using the Lie algebroid generalized tangent bundle of a Lie
algebroid, we obtain a new solution for the \textbf{Weinstein's Problem},
different by the Martinez's solution.

Finally, we obtain that the\ integral curves of the canonical $\left( \rho
,\eta \right) $-semispray associated to Lagrange mechanical $\left( \rho
,\eta \right) $-system $\left( \left( E,\pi ,M\right) ,F_{e},L\right) $\ and
from locally invertible\emph{\ }$\mathbf{B}^{\mathbf{v}}$-morphism $\left(
g,h\right) $\ are the $\left( g,h\right) $-lifts solutions for the equations
of Euler-Lagrange type $\left( 9.10\right) $.

Using our theory, we obtain the following

\textbf{Theorem }\emph{If }$F$\emph{\ is a Finsler fundamental function,
then the\ geodesics on the manifold }$M$\emph{\ are the curves such that the
components of their }$\left( g,h\right) $\emph{-lifts are\ solutions for the
equations of Euler-Lagrange type }$\left( 9.10\right) .$

As any Lie algebroid can be regarded as a particularly generalized Lie
algebroid, it is natural to propose to extend the study of the Finsler
geometry from the usual Lie algebroid $\left( \left( TM,\tau _{M},M\right) ,%
\left[ ,\right] _{TM},\left( Id_{TM},Id_{M}\right) \right) ,$ to an
arbitrary (generalized) Lie algebroid $\left( \left( E,\pi ,M\right) ,\left[
,\right] _{E,h},\left( \rho ,\eta \right) \right) .$

\section{Preliminaries}

Let$\mathbf{~Vect},$ $\mathbf{Liealg},~\mathbf{Mod}$\textbf{,} $\mathbf{Man}$
and $\mathbf{B}^{\mathbf{v}}$ be the category of real vector spaces, Lie
algebras, modules, manifolds and vector bundles respectively.

We know that if $\left( E,\pi ,M\right) \in \left\vert \mathbf{B}^{\mathbf{v}%
}\right\vert $ so that $M$ is paracompact and if $A\subseteq M$ is closed,
then for any section $u$ over $A$ it exists $\tilde{u}\in $ $\Gamma \left(
E,\pi ,M\right) $ so that $\tilde{u}_{|A}=u.$ In the following, we consider
only vector bundles with paracompact base.

Aditionally, if $\left( E,\pi ,M\right) \in \left\vert \mathbf{B}^{\mathbf{v}%
}\right\vert ,$ $\Gamma \left( E,\pi ,M\right) =\left\{ u\in \mathbf{Man}%
\left( M,E\right) :u\circ \pi =Id_{M}\right\} $ and $\mathcal{F}\left(
M\right) =\mathbf{Man}\left( M,\mathbb{R}\right) ,$ then $\left( \Gamma
\left( E,\pi ,M\right) ,+,\cdot \right) $ is a $\mathcal{F}\left( M\right) $%
-module. If \ $\left( \varphi ,\varphi _{0}\right) \in \mathbf{B}^{\mathbf{v}%
}\left( \left( E,\pi ,M\right) ,\left( E^{\prime },\pi ^{\prime },M^{\prime
}\right) \right) $ such that $\varphi _{0}\in Iso_{\mathbf{Man}}\left(
M,M^{\prime }\right) ,$ then, using the operation
\begin{equation*}
\begin{array}{ccc}
\mathcal{F}\left( M\right) \times \Gamma \left( E^{\prime },\pi ^{\prime
},M^{\prime }\right) & ^{\underrightarrow{~\ \ \cdot ~\ \ }} & \Gamma \left(
E^{\prime },\pi ^{\prime },M^{\prime }\right) \\
\left( f,u^{\prime }\right) & \longmapsto & f\circ \varphi _{0}^{-1}\cdot
u^{\prime }%
\end{array}%
\end{equation*}%
it results that $\left( \Gamma \left( E^{\prime },\pi ^{\prime },M^{\prime
}\right) ,+,\cdot \right) $ is a $\mathcal{F}\left( M\right) $-module and we
obtain the $\mathbf{Mod}$-morphism%
\begin{equation*}
\begin{array}{ccc}
\Gamma \left( E,\pi ,M\right) & ^{\underrightarrow{~\ \ \Gamma \left(
\varphi ,\varphi _{0}\right) ~\ \ }} & \Gamma \left( E^{\prime },\pi
^{\prime },M^{\prime }\right) \\
u & \longmapsto & \Gamma \left( \varphi ,\varphi _{0}\right) u%
\end{array}%
\end{equation*}%
defined by
\begin{equation*}
\begin{array}{c}
\Gamma \left( \varphi ,\varphi _{0}\right) u\left( y\right) =\varphi \left(
u_{\varphi _{0}^{-1}\left( y\right) }\right) ,%
\end{array}%
\end{equation*}%
for any $y\in M^{\prime }.$

Let $M,N\in \left\vert \mathbf{Man}\right\vert ,$ $h\in Iso_{\mathbf{Man}%
}\left( M,N\right) $ and $\eta \in Iso_{\mathbf{Man}}\left( N,M\right) $.

We know (see $\left[ 2,3\right] $) that if $\left( F,\nu ,N\right) \in
\left\vert \mathbf{B}^{\mathbf{v}}\right\vert $ so that there exists
\begin{equation*}
\begin{array}{c}
\left( \rho ,\eta \right) \in \mathbf{B}^{\mathbf{v}}\left( \left( F,\nu
,N\right) ,\left( TM,\tau _{M},M\right) \right)%
\end{array}%
\end{equation*}%
and an operation
\begin{equation*}
\begin{array}{ccc}
\Gamma \left( F,\nu ,N\right) \times \Gamma \left( F,\nu ,N\right) & ^{%
\underrightarrow{\left[ ,\right] _{F,h}}} & \Gamma \left( F,\nu ,N\right) \\
\left( u,v\right) & \longmapsto & \left[ u,v\right] _{F,h}%
\end{array}%
\end{equation*}%
with the following properties:\bigskip

\noindent $\qquad GLA_{1}$. the equality holds good
\begin{equation*}
\begin{array}{c}
\left[ u,f\cdot v\right] _{F,h}=f\left[ u,v\right] _{F,h}+\Gamma \left(
Th\circ \rho ,h\circ \eta \right) \left( u\right) f\cdot v,%
\end{array}%
\end{equation*}%
\qquad \quad\ \ for all $u,v\in \Gamma \left( F,\nu ,N\right) $ and $f\in
\mathcal{F}\left( N\right) .$

\medskip $GLA_{2}$. the $4$-tuple $\left( \Gamma \left( F,\nu ,N\right)
,+,\cdot ,\left[ ,\right] _{F,h}\right) $ is a Lie $\mathcal{F}\left(
N\right) $-algebra,

$GLA_{3}$. the $\mathbf{Mod}$-morphism $\Gamma \left( Th\circ \rho ,h\circ
\eta \right) $ is a $\mathbf{LieAlg}$-morphism of
\begin{equation*}
\left( \Gamma \left( F,\nu ,N\right) ,+,\cdot ,\left[ ,\right] _{F,h}\right)
\end{equation*}%
source and
\begin{equation*}
\left( \Gamma \left( TN,\tau _{N},N\right) ,+,\cdot ,\left[ ,\right]
_{TN}\right)
\end{equation*}%
target, \medskip \noindent then the triple $\left( \left( F,\nu ,N\right) ,%
\left[ ,\right] _{F,h},\left( \rho ,\eta \right) \right) $ is called
generalized Lie algebroid.\emph{\ }

In particular, if $h=Id_{M}=\eta ,$ then we obtain the definition of the Lie
algebroid.

We can discuss about \emph{the category }$\mathbf{GLA}$\emph{\ of
generalized Lie algebroids }$\left[ 2\right] $.

Examples of objects of this category are presented in the paper $\left[ 1%
\right] .$

Let $\left( \left( F,\nu ,N\right) ,\left[ ,\right] _{F,h},\left( \rho ,\eta
\right) \right) $ be an object of the category $\mathbf{GLA}$.

\begin{itemize}
\item Locally, for any $\alpha ,\beta \in \overline{1,p},$ we set $\left[
t_{\alpha },t_{\beta }\right] _{F,h}=L_{\alpha \beta }^{\gamma }t_{\gamma }.$
We easily obtain that $L_{\alpha \beta }^{\gamma }=-L_{\beta \alpha
}^{\gamma },~$for any $\alpha ,\beta ,\gamma \in \overline{1,p}.$
\end{itemize}

The real local functions $L_{\alpha \beta }^{\gamma },~\alpha ,\beta ,\gamma
\in \overline{1,p}$ will be called the \emph{structure functions of the
generalized Lie algebroid }$\left( \left( F,\nu ,N\right) ,\left[ ,\right]
_{F,h},\left( \rho ,\eta \right) \right) .$

\begin{itemize}
\item We assume the following diagrams:%
\begin{equation*}
\begin{array}[b]{ccccc}
F & ^{\underrightarrow{~\ \ \ \rho ~\ \ }} & TM & ^{\underrightarrow{~\ \ \
Th~\ \ }} & TN \\
~\downarrow \nu &  & ~\ \ \ \downarrow \tau _{M} &  & ~\ \ \ \downarrow \tau
_{N} \\
N & ^{\underrightarrow{~\ \ \ \eta ~\ \ }} & M & ^{\underrightarrow{~\ \ \
h~\ \ }} & N \\
&  &  &  &  \\
\left( \chi ^{\tilde{\imath}},z^{\alpha }\right) &  & \left(
x^{i},y^{i}\right) &  & \left( \chi ^{\tilde{\imath}},z^{\tilde{\imath}%
}\right)%
\end{array}%
\end{equation*}

where $i,\tilde{\imath}\in \overline{1,m}$ and $\alpha \in \overline{1,p}.$

If%
\begin{equation*}
\left( \chi ^{\tilde{\imath}},z^{\alpha }\right) \longrightarrow \left( \chi
^{\tilde{\imath}\prime }\left( \chi ^{\tilde{\imath}}\right) ,z^{\alpha
\prime }\left( \chi ^{\tilde{\imath}},z^{\alpha }\right) \right) ,
\end{equation*}%
\begin{equation*}
\left( x^{i},y^{i}\right) \longrightarrow \left( x^{i%
{\acute{}}%
}\left( x^{i}\right) ,y^{i%
{\acute{}}%
}\left( x^{i},y^{i}\right) \right)
\end{equation*}%
and
\begin{equation*}
\left( \chi ^{\tilde{\imath}},z^{\tilde{\imath}}\right) \longrightarrow
\left( \chi ^{\tilde{\imath}\prime }\left( \chi ^{\tilde{\imath}}\right) ,z^{%
\tilde{\imath}\prime }\left( \chi ^{\tilde{\imath}},z^{\tilde{\imath}%
}\right) \right) ,
\end{equation*}%
then
\begin{equation*}
\begin{array}[b]{c}
z^{\alpha
{\acute{}}%
}=\Lambda _{\alpha }^{\alpha
{\acute{}}%
}z^{\alpha }%
\end{array}%
,
\end{equation*}%
\begin{equation*}
\begin{array}[b]{c}
y^{i%
{\acute{}}%
}=\frac{\partial x^{i%
{\acute{}}%
}}{\partial x^{i}}y^{i}%
\end{array}%
\end{equation*}%
and
\begin{equation*}
\begin{array}{c}
z^{\tilde{\imath}\prime }=\frac{\partial \chi ^{\tilde{\imath}\prime }}{%
\partial \chi ^{\tilde{\imath}}}z^{\tilde{\imath}}.%
\end{array}%
\end{equation*}

\item We assume that $\left( \theta ,\mu \right) \overset{put}{=}\left(
Th\circ \rho ,h\circ \eta \right) $. If $z^{\alpha }t_{\alpha }\in \Gamma
\left( F,\nu ,N\right) $ is arbitrary, then
\begin{equation*}
\begin{array}[t]{l}
\begin{array}{c}
\Gamma \left( Th\circ \rho ,h\circ \eta \right) \left( z^{\alpha }t_{\alpha
}\right) f\left( h\circ \eta \left( \varkappa \right) \right) =\vspace*{1mm}
\\
=\left( \theta _{\alpha }^{\tilde{\imath}}z^{\alpha }\frac{\partial f}{%
\partial \varkappa ^{\tilde{\imath}}}\right) \left( h\circ \eta \left(
\varkappa \right) \right) =\left( \left( \rho _{\alpha }^{i}\circ h\right)
\left( z^{\alpha }\circ h\right) \frac{\partial f\circ h}{\partial x^{i}}%
\right) \left( \eta \left( \varkappa \right) \right) ,%
\end{array}%
\end{array}%
\leqno(2.1)
\end{equation*}%
for any $f\in \mathcal{F}\left( N\right) $ and $\varkappa \in N.$
\end{itemize}

The coefficients $\rho _{\alpha }^{i}$ respectively $\theta _{\alpha }^{%
\tilde{\imath}}$ change to $\rho _{\alpha
{\acute{}}%
}^{i%
{\acute{}}%
}$ respectively $\theta _{\alpha
{\acute{}}%
}^{\tilde{\imath}%
{\acute{}}%
}$ according to the rule:
\begin{equation*}
\begin{array}{c}
\rho _{\alpha
{\acute{}}%
}^{i%
{\acute{}}%
}=\Lambda _{\alpha
{\acute{}}%
}^{\alpha }\rho _{\alpha }^{i}\displaystyle\frac{\partial x^{i%
{\acute{}}%
}}{\partial x^{i}},%
\end{array}%
\leqno(2.2)
\end{equation*}%
respectively%
\begin{equation*}
\begin{array}{c}
\theta _{\alpha
{\acute{}}%
}^{\tilde{\imath}%
{\acute{}}%
}=\Lambda _{\alpha
{\acute{}}%
}^{\alpha }\theta _{\alpha }^{\tilde{\imath}}\displaystyle\frac{\partial
\varkappa ^{\tilde{\imath}%
{\acute{}}%
}}{\partial \varkappa ^{\tilde{\imath}}},%
\end{array}%
\leqno(2.3)
\end{equation*}%
where
\begin{equation*}
\left\Vert \Lambda _{\alpha
{\acute{}}%
}^{\alpha }\right\Vert =\left\Vert \Lambda _{\alpha }^{\alpha
{\acute{}}%
}\right\Vert ^{-1}.
\end{equation*}

\emph{Remark 2.1 } The following equalities hold good:%
\begin{equation*}
\begin{array}{c}
\displaystyle\rho _{\alpha }^{i}\circ h\frac{\partial f\circ h}{\partial
x^{i}}=\left( \theta _{\alpha }^{\tilde{\imath}}\frac{\partial f}{\partial
\varkappa ^{\tilde{\imath}}}\right) \circ h,\forall f\in \mathcal{F}\left(
N\right) .%
\end{array}%
\leqno(2.4)
\end{equation*}%
\emph{and }%
\begin{equation*}
\begin{array}{c}
\displaystyle\left( L_{\alpha \beta }^{\gamma }\circ h\right) \left( \rho
_{\gamma }^{k}\circ h\right) =\left( \rho _{\alpha }^{i}\circ h\right) \frac{%
\partial \left( \rho _{\beta }^{k}\circ h\right) }{\partial x^{i}}-\left(
\rho _{\beta }^{j}\circ h\right) \frac{\partial \left( \rho _{\alpha
}^{k}\circ h\right) }{\partial x^{j}}.%
\end{array}%
\leqno(2.5)
\end{equation*}

We have the $\mathbf{B}^{\mathbf{v}}$-morphism%
\begin{equation*}
\begin{array}{ccc}
~\ \ \ \ \ \ \ \ \ \ \ \ \ \pi ^{\ast }\left( h^{\ast }F\right) &
\hookrightarrow & F \\
\pi ^{\ast }\left( h^{\ast }\nu \right) \downarrow &  & ~\downarrow \nu \\
~\ \ \ \ \ \ \ \ \ \ \ \ M & ^{\underrightarrow{~\ \ h\circ \pi ~\ \ }} & N%
\end{array}%
\leqno(2.6)
\end{equation*}

Let $\Big(\overset{\pi ^{\ast }\left( h^{\ast }F\right) }{\rho },Id_{E}\Big)$
be the\emph{\ }$\mathbf{B}^{\mathbf{v}}$-morphism of\emph{\ }$\left( \pi
^{\ast }\left( h^{\ast }F\right) ,\pi ^{\ast }\left( h^{\ast }\nu \right)
,E\right) $\ source and $\left( TE,\tau _{E},E\right) $\ target, where%
\begin{equation*}
\begin{array}{rcl}
\pi ^{\ast }\left( h^{\ast }F\right) & ^{\underrightarrow{\overset{\pi
^{\ast }\left( h^{\ast }F\right) }{\rho }}} & TE \\
\displaystyle Z^{\alpha }T_{\alpha }\left( u_{x}\right) & \longmapsto & %
\displaystyle\left( Z^{\alpha }\cdot \rho _{\alpha }^{i}\circ h\circ \pi
\right) \frac{\partial }{\partial x^{i}}\left( u_{x}\right)%
\end{array}%
\leqno(2.7)
\end{equation*}

Using the operation
\begin{equation*}
\begin{array}{ccc}
\Gamma \left( \pi ^{\ast }\left( h^{\ast }F\right) ,\pi ^{\ast }\left(
h^{\ast }\nu \right) ,E\right) ^{2} & ^{\underrightarrow{~\ \ \left[ ,\right]
_{\pi ^{\ast }\left( h^{\ast }F\right) }~\ \ }} & \Gamma \left( \pi ^{\ast
}\left( h^{\ast }F\right) ,\pi ^{\ast }\left( h^{\ast }\nu \right) ,E\right)%
\end{array}%
\end{equation*}%
defined by%
\begin{equation*}
\begin{array}{ll}
\left[ T_{\alpha },T_{\beta }\right] _{\pi ^{\ast }\left( h^{\ast }F\right) }
& =\left( L_{\alpha \beta }^{\gamma }\circ h\circ \pi \right) T_{\gamma },%
\vspace*{1mm} \\
\left[ T_{\alpha },fT_{\beta }\right] _{\pi ^{\ast }\left( h^{\ast }F\right)
} & \displaystyle=f\left( L_{\alpha \beta }^{\gamma }\circ h\circ \pi
\right) T_{\gamma }+\left( \rho _{\alpha }^{i}\circ h\circ \pi \right) \frac{%
\partial f}{\partial x^{i}}T_{\beta },\vspace*{1mm} \\
\left[ fT_{\alpha },T_{\beta }\right] _{\pi ^{\ast }\left( h^{\ast }F\right)
} & =-\left[ T_{\beta },fT_{\alpha }\right] _{\pi ^{\ast }\left( h^{\ast
}F\right) },%
\end{array}%
\leqno(2.8)
\end{equation*}%
for any $f\in \mathcal{F}\left( E\right) ,$ it results that
\begin{equation*}
\begin{array}{c}
\left( \left( \pi ^{\ast }\left( h^{\ast }F\right) ,\pi ^{\ast }\left(
h^{\ast }\nu \right) ,E\right) ,\left[ ,\right] _{\pi ^{\ast }\left( h^{\ast
}F\right) },\left( \overset{\pi ^{\ast }\left( h^{\ast }F\right) }{\rho }%
,Id_{E}\right) \right)%
\end{array}%
\end{equation*}%
is a Lie algebroid.

\section{Natural and adapted basis}

We consider the following diagram:
\begin{equation*}
\begin{array}{c}
\xymatrix{E\ar[d]_\pi&\left( F,\left[ ,\right] _{F,h},\left( \rho ,\eta
\right) \right)\ar[d]^\nu\\ M\ar[r]^h&N}%
\end{array}%
\leqno(3.1)
\end{equation*}%
where $\left( E,\pi ,M\right) $ is a vector bundle and $\left( \left( F,\nu
,N\right) ,\left[ ,\right] _{F,h},\left( \rho ,\eta \right) \right) $ is a
generalized Lie algebroid.

We take $\left( x^{i},y^{a}\right) $ as canonical local coordinates on $%
\left( E,\pi ,M\right) ,$ where $i\in \overline{1,m}$ and $a\in \overline{1,r%
}.$ Let
\begin{equation*}
\left( x^{i},y^{a}\right) \longrightarrow \left( x^{i%
{\acute{}}%
}\left( x^{i}\right) ,y^{a%
{\acute{}}%
}\left( x^{i},y^{a}\right) \right)
\end{equation*}%
be a change of coordinates on $\left( E,\pi ,M\right) $. Then the
coordinates $y^{a}$ change to $y^{a%
{\acute{}}%
}$ by the rule:
\begin{equation*}
\begin{array}{c}
y^{a%
{\acute{}}%
}=M_{a}^{a%
{\acute{}}%
}y^{a}.%
\end{array}%
\leqno(3.2)
\end{equation*}

Let
\begin{equation*}
\begin{array}[b]{c}
\left( \frac{\partial }{\partial x^{i}},\frac{\partial }{\partial y^{a}}%
\right) \overset{put}{=}\left( \partial _{i},\dot{\partial}_{a}\right)%
\end{array}%
\leqno(3.3)
\end{equation*}%
be the natural base of the Lie algebroid $\left( \left( TE,\tau
_{E},E\right) ,\left[ ,\right] _{TE},\left( Id_{TE},Id_{E}\right) \right) .$

For any sections%
\begin{equation*}
\begin{array}{c}
Z^{\alpha }T_{\alpha }\in \Gamma \left( \pi ^{\ast }\left( h^{\ast }F\right)
,\pi ^{\ast }\left( h^{\ast }F\right) ,E\right)%
\end{array}%
\end{equation*}%
and%
\begin{equation*}
\begin{array}{c}
Y^{a}\dot{\partial}_{a}\in \Gamma \left( VTE,\tau _{E},E\right)%
\end{array}%
\end{equation*}%
we obtain the section%
\begin{equation*}
\begin{array}{c}
Z^{\alpha }\displaystyle\tilde{\partial}_{\alpha }+Y^{a}\overset{\cdot }{%
\tilde{\partial}}_{a}=:Z^{\alpha }\left( T_{\alpha }\oplus \left( \rho
_{\alpha }^{i}\circ h\circ \pi \right) \partial _{i}\right) +Y^{a}\left(
0_{\pi ^{\ast }\left( h^{\ast }F\right) }\oplus \dot{\partial}_{a}\right)
\vspace*{1mm} \\
=\displaystyle Z^{\alpha }T_{\alpha }\oplus \left( Z^{\alpha }\left( \rho
_{\alpha }^{i}\circ h\circ \pi \right) \partial _{i}+Y^{a}\dot{\partial}%
_{a}\right) \in \Gamma \left( \pi ^{\ast }\left( h^{\ast }F\right) \oplus TE,%
\overset{\oplus }{\pi },E\right) .%
\end{array}%
\end{equation*}

Since we have
\begin{equation*}
\begin{array}{c}
Z^{\alpha }\displaystyle\tilde{\partial}_{\alpha }+Y^{a}\overset{\cdot }{%
\tilde{\partial}}_{a}=0 \\
\Updownarrow \\
Z^{\alpha }T_{\alpha }=0~\wedge Z^{\alpha }\left( \rho _{\alpha }^{i}\circ
h\circ \pi \right) \displaystyle\partial _{i}+Y^{a}\dot{\partial}_{a}=0,%
\end{array}%
\end{equation*}%
it implies $Z^{\alpha }=0,~\alpha \in \overline{1,p}$ and $Y^{a}=0,~a\in
\overline{1,r}.$

Therefore, the sections $\displaystyle\tilde{\partial}_{1},...,\tilde{%
\partial}_{p},\overset{\cdot }{\tilde{\partial}}_{1},...,\overset{\cdot }{%
\tilde{\partial}}_{r}$ are linearly independent.\smallskip

We consider the vector subbundle $\left( \left( \rho ,\eta \right) TE,\left(
\rho ,\eta \right) \tau _{E},E\right) $ of the vector bundle\break $\left(
\pi ^{\ast }\left( h^{\ast }F\right) \oplus TE,\overset{\oplus }{\pi }%
,E\right) ,$ for which the $\mathcal{F}\left( E\right) $-module of sections
is the $\mathcal{F}\left( E\right) $-submodule of $\left( \Gamma \left( \pi
^{\ast }\left( h^{\ast }F\right) \oplus TE,\overset{\oplus }{\pi },E\right)
,+,\cdot \right) ,$ generated by the set of sections $\left( \tilde{\partial}%
_{\alpha },\overset{\cdot }{\tilde{\partial}}_{a}\right) $ which is called
the \emph{natural }$\left( \rho ,\eta \right) $\emph{-base.}

The matrix of coordinate transformation on $\left( \left( \rho ,\eta \right)
TE,\left( \rho ,\eta \right) \tau _{E},E\right) $ at a change of fibred
charts is
\begin{equation*}
\left\Vert
\begin{array}{cc}
\Lambda _{\alpha }^{\alpha
{\acute{}}%
}\circ h\circ \pi & 0\vspace*{1mm} \\
\left( \rho _{a}^{i}\circ h\circ \pi \right) \displaystyle\frac{\partial
M_{b}^{a%
{\acute{}}%
}\circ \pi }{\partial x_{i}}y^{b} & M_{a}^{a%
{\acute{}}%
}\circ \pi%
\end{array}%
\right\Vert .\leqno(3.4)
\end{equation*}

We have the following

\textbf{Theorem 3.1 }\emph{Let} $\left( \tilde{\rho},Id_{E}\right) $\ \emph{%
be the} $\mathbf{B}^{\mathbf{v}}$\emph{-morphism of }$\left( \left( \rho
,\eta \right) TE,\left( \rho ,\eta \right) \tau _{E},E\right) $\ \emph{%
source and }$\left( TE,\tau _{E},E\right) $\ \emph{target, where}
\begin{equation*}
\begin{array}{rcl}
\left( \rho ,\eta \right) TE\!\!\! & \!\!^{\underrightarrow{\tilde{\ \ \rho
\ \ }}}\!\!\! & \!\!TE\vspace*{2mm} \\
\left( Z^{\alpha }\displaystyle\tilde{\partial}_{\alpha }+Y^{a}\overset{%
\cdot }{\tilde{\partial}}_{a}\right) \!(u_{x})\!\!\!\! & \!\!\longmapsto
\!\!\! & \!\!\left( \!Z^{\alpha }\!\left( \rho _{\alpha }^{i}{\circ }h{\circ
}\pi \!\right) \!\partial _{i}{+}Y^{a}\dot{\partial}_{a}\right)
\!(u_{x})\!\!.%
\end{array}%
\leqno(3.5)
\end{equation*}

\emph{Using the operation}
\begin{equation*}
\begin{array}{ccc}
\Gamma \left( \left( \rho ,\eta \right) TE,\left( \rho ,\eta \right) \tau
_{E},E\right) ^{2} & ^{\underrightarrow{~\ \ \left[ ,\right] _{\left( \rho
,\eta \right) TE}~\ \ }} & \Gamma \left( \left( \rho ,\eta \right) TE,\left(
\rho ,\eta \right) \tau _{E},E\right)%
\end{array}%
\end{equation*}%
\emph{defined by}%
\begin{equation*}
\begin{array}{l}
\left[ \left( Z_{1}^{\alpha }\displaystyle\tilde{\partial}_{\alpha
}+Y_{1}^{a}\overset{\cdot }{\tilde{\partial}}_{a}\right) ,\left(
Z_{2}^{\beta }\displaystyle\tilde{\partial}_{\beta }+Y_{2}^{b}\displaystyle%
\overset{\cdot }{\tilde{\partial}}_{b}\right) \right] _{\left( \rho ,\eta
\right) TE}\vspace*{1mm} \\
\displaystyle=\left[ Z_{1}^{\alpha }T_{a},Z_{2}^{\beta }T_{\beta }\right]
_{\pi ^{\ast }\left( h^{\ast }F\right) }\oplus \left[ \left( \rho _{\alpha
}^{i}\circ h\circ \pi \right) Z_{1}^{\alpha }\partial _{i}+Y_{1}^{a}\dot{%
\partial}_{a},\right. \vspace*{1mm} \\
\hfill \displaystyle\left. \left( \rho _{\beta }^{j}\circ h\circ \pi \right)
Z_{2}^{\beta }\partial _{j}+Y_{2}^{b}\dot{\partial}_{b}\right] _{TE},%
\end{array}%
\leqno(3.6)
\end{equation*}%
\emph{for any} $\left( Z_{1}^{\alpha }\displaystyle\tilde{\partial}_{\alpha
}+Y_{1}^{a}\overset{\cdot }{\tilde{\partial}}_{a}\right) $\emph{\ and }$%
\left( Z_{2}^{\beta }\displaystyle\tilde{\partial}_{\beta }+Y_{2}^{b}\overset%
{\cdot }{\tilde{\partial}}_{b}\right) ,$ \emph{we obtain that the couple }%
\begin{equation*}
\left( \left[ ,\right] _{\left( \rho ,\eta \right) TE},\left( \tilde{\rho}%
,Id_{E}\right) \right)
\end{equation*}%
\emph{\ is a Lie algebroid structure for the vector bundle }$\left( \left(
\rho ,\eta \right) TE,\left( \rho ,\eta \right) \tau _{E},E\right) .$

The Lie algebroid
\begin{equation*}
\begin{array}{c}
\left( \left( \left( \rho ,\eta \right) TE,\left( \rho ,\eta \right) \tau
_{E},E\right) ,\left[ ,\right] _{\left( \rho ,\eta \right) TE},\left( \tilde{%
\rho},Id_{E}\right) \right)%
\end{array}%
,
\end{equation*}%
is called the \emph{Lie algebroid generalized tangent bundle.}

\textit{Remark 3.1 }The following equalities hold good:%
\begin{equation*}
\begin{array}{c}
\begin{array}[b]{cl}
\left[ \tilde{\partial}_{\alpha },\tilde{\partial}_{\beta }\right] _{\left(
\rho ,\eta \right) TE} & =L_{\alpha \beta }^{\gamma }\circ h\circ \pi \cdot
\tilde{\partial}_{\gamma } \\
\left[ \tilde{\partial}_{\alpha },\tilde{\partial}_{b}\right] _{\left( \rho
,\eta \right) TE} & =0_{\left( \rho ,\eta \right) TE} \\
\left[ \tilde{\partial}_{a},\tilde{\partial}_{b}\right] _{\left( \rho ,\eta
\right) TE} & =0_{\left( \rho ,\eta \right) TE}%
\end{array}%
\end{array}%
\leqno(3.7)
\end{equation*}

We consider the $\mathbf{B}^{\mathbf{v}}$-morphism $\left( \left( \rho ,\eta
\right) \pi !,Id_{E}\right) $ given by the commutative diagram%
\begin{equation*}
\begin{array}{c}
\xymatrix{\left( \rho ,\eta \right) TE\ar[r]^{( \rho ,\eta ) \pi
!}\ar[d]_{(\rho,\eta)\tau_E}&\pi ^{\ast }\left( h^{\ast
}F\right)\ar[d]^{pr_1} \\ E\ar[r]^{id_{E}}& E}%
\end{array}%
\leqno(3.8)
\end{equation*}

This is defined as:%
\begin{equation*}
\begin{array}{c}
\left( \rho ,\eta \right) \pi !\left( \left( Z^{\alpha }\tilde{\partial}%
_{\alpha }+Y^{a}\overset{\cdot }{\tilde{\partial}}_{a}\right) \left(
u_{x}\right) \right) =\left( Z^{\alpha }T_{\alpha }\right) \left(
u_{x}\right) ,%
\end{array}%
\leqno(3.9)
\end{equation*}%
for any $\displaystyle\left( Z^{\alpha }\tilde{\partial}_{\alpha }+Y^{a}%
\displaystyle\overset{\cdot }{\tilde{\partial}}_{a}\right) \in \Gamma \left(
\left( \rho ,\eta \right) TE,\left( \rho ,\eta \right) \tau _{E},E\right) .$%
\medskip

Using the $\mathbf{B}^{\mathbf{v}}$-morphisms $\left( 2.6\right) $ and $%
\left( 3.7\right) $ we obtain the \emph{tangent }$\left( \rho ,\eta \right) $%
\emph{-application }$\left( \left( \rho ,\eta \right) T\pi ,h\circ \pi
\right) $ of $\left( \left( \rho ,\eta \right) TE,\left( \rho ,\eta \right)
\tau _{E},E\right) $ source and $\left( F,\nu ,N\right) $ target.

\textbf{Definition 3.1} The kernel of the tangent $\left( \rho ,\eta \right)
$-application\ is written
\begin{equation*}
\left( V\left( \rho ,\eta \right) TE,\left( \rho ,\eta \right) \tau
_{E},E\right)
\end{equation*}%
and it is called \emph{the vertical subbundle}.\bigskip

We remark that the set $\left\{ \displaystyle\overset{\cdot }{\tilde{\partial%
}}_{a},~a\in \overline{1,r}\right\} $ is a base of the $\mathcal{F}\left(
E\right) $-module
\begin{equation*}
\left( \Gamma \left( V\left( \rho ,\eta \right) TE,\left( \rho ,\eta \right)
\tau _{E},E\right) ,+,\cdot \right) .
\end{equation*}

\textbf{Proposition 3.1} \emph{The short sequence of vector bundles}%
\begin{equation*}
\begin{array}{c}
\xymatrix{0\ar@{^(->}[r]^i\ar[d]&V(\rho,\eta)TE\ar[d]\ar@{^(->}[r]^i&(\rho,%
\eta)TE\ar[r]^{(\rho,\eta)\pi!}\ar[d]&\pi ^{\ast }\left( h^{\ast
}F\right)\ar[r]\ar[d]&0\ar[d]\\
E\ar[r]^{Id_E}&E\ar[r]^{Id_E}&E\ar[r]^{Id_E}&E\ar[r]^{Id_E}&E}%
\end{array}%
\leqno(3.10)
\end{equation*}%
\emph{is exact.}

Let $\left( \rho ,\eta \right) \Gamma $ be a $\left( \rho ,\eta \right) $%
-connection for the vector bundle$\left( E,\pi ,M\right) ,$ i. e. \textit{a }%
$\mathbf{Man}$-morphism $\left( \rho ,\eta \right) \Gamma $ of $\left( \rho
,\eta \right) TE$ source and $V\left( \rho ,\eta \right) TE$ target defined
by%
\begin{equation*}
\begin{array}{c}
\left( \rho ,\eta \right) \Gamma \left( Z^{\alpha }\tilde{\partial}_{\alpha
}+Y^{a}\overset{\cdot }{\tilde{\partial}}_{a}\right) \left( u_{x}\right)
=\left( Y^{a}+\left( \rho ,\eta \right) \Gamma _{\alpha }^{a}Z^{\alpha
}\right) \overset{\cdot }{\tilde{\partial}}_{a}\left( u_{x}\right) ,%
\end{array}%
\leqno(3.11)
\end{equation*}%
so that the $\mathbf{B}^{\mathbf{v}}$-morphism $\left( \left( \rho ,\eta
\right) \Gamma ,Id_{E}\right) $ is a split to the left in the previous exact
sequence. Its components satisfy the law of transformation\emph{\ }%
\begin{equation*}
\begin{array}{c}
\left( \rho ,\eta \right) \Gamma _{\gamma
{\acute{}}%
}^{a%
{\acute{}}%
}{=}M_{a}^{a%
{\acute{}}%
}{\circ }\pi \!\!\left[ \rho _{\gamma }^{i}{\circ }\left( h{\circ }\pi
\right) \!\frac{\partial M_{b%
{\acute{}}%
}^{a}\circ \pi }{\partial x^{i}}y^{b%
{\acute{}}%
}{+}\left( \rho ,\eta \right) \!\Gamma _{\gamma }^{a}\right] \!\!\Lambda
_{\gamma
{\acute{}}%
}^{\gamma }{\circ }\left( h{\circ }\pi \right) .%
\end{array}%
\leqno(3.12)
\end{equation*}

The kernel of the $\mathbf{B}^{\mathbf{v}}$-morphism $\left( \left( \rho
,\eta \right) \Gamma ,Id_{E}\right) $\ is written $\left( H\left( \rho ,\eta
\right) TE,\left( \rho ,\eta \right) \tau _{E},E\right) $ and is called the
\emph{horizontal vector subbundle}.

We remark that the horizontal and the vertical vector subbundles are
interior differential systems of the Lie algebroid generalized tangent
bundle. (see $\left[ 4\right] $)

We put the problem of finding a base for the $\mathcal{F}\left( E\right) $%
-module
\begin{equation*}
\left( \Gamma \left( H\left( \rho ,\eta \right) TE,\left( \rho ,\eta \right)
\tau _{E},E\right) ,+,\cdot \right)
\end{equation*}%
of the type\textbf{\ }
\begin{equation*}
\begin{array}[t]{l}
\tilde{\delta}_{\alpha }=Z_{\alpha }^{\beta }\tilde{\partial}_{\alpha
}+Y_{\alpha }^{a}\overset{\cdot }{\tilde{\partial}}_{a},\alpha \in \overline{%
1,r}%
\end{array}%
\end{equation*}%
which satisfies the following conditions:
\begin{equation*}
\begin{array}{rcl}
\displaystyle\Gamma \left( \left( \rho ,\eta \right) \pi !,Id_{E}\right)
\left( \tilde{\delta}_{\alpha }\right) & = & T_{\alpha }\vspace*{2mm}, \\
\displaystyle\Gamma \left( \left( \rho ,\eta \right) \Gamma ,Id_{E}\right)
\left( \tilde{\delta}_{\alpha }\right) & = & 0.%
\end{array}%
\leqno(3.13)
\end{equation*}

Then we obtain the sections
\begin{equation*}
\begin{array}[t]{l}
\frac{\delta }{\delta \tilde{z}^{\alpha }}=\tilde{\partial}_{\alpha }-\left(
\rho ,\eta \right) \Gamma _{\alpha }^{a}\overset{\cdot }{\tilde{\partial}}%
_{a}=T_{\alpha }\oplus \left( \left( \rho _{\alpha }^{i}\circ h\circ \pi
\right) \partial _{i}-\left( \rho ,\eta \right) \Gamma _{\alpha }^{a}\dot{%
\partial}_{a}\right) .%
\end{array}%
\leqno(3.14)
\end{equation*}%
such that their law of change is a tensorial law under a change of vector
fiber charts.

The base $\left( \tilde{\delta}_{\alpha },\overset{\cdot }{\tilde{\partial}}%
_{a}\right) $ will be called the \emph{adapted }$\left( \rho ,\eta \right) $%
\emph{-base.}

\textit{Remark 3.2 }The following equality holds good%
\begin{equation*}
\begin{array}{l}
\Gamma \left( \tilde{\rho},Id_{E}\right) \left( \tilde{\delta}_{\alpha
}\right) =\left( \rho _{\alpha }^{i}\circ h\circ \pi \right) \partial
_{i}-\left( \rho ,\eta \right) \Gamma _{\alpha }^{a}\dot{\partial}_{a}.%
\end{array}%
\leqno(3.15)
\end{equation*}

Moreover, if $\left( \rho ,\eta \right) \Gamma $ is the $\left( \rho ,\eta
\right) $-connection associated to a connection $\Gamma $ (see $\left[ 1%
\right] $), then we obtain
\begin{equation*}
\begin{array}{l}
\Gamma \left( \tilde{\rho},Id_{E}\right) \left( \tilde{\delta}_{\alpha
}\right) =\left( \rho _{\alpha }^{i}\circ h\circ \pi \right) \delta _{i},%
\end{array}%
\leqno(3.16)
\end{equation*}%
where $\left( \delta _{i},\dot{\partial}_{a}\right) $ is the adapted base
for the $\mathcal{F}\left( E\right) $-module $\left( \Gamma \left( TE,\tau
_{E},E\right) ,+,\cdot \right) .$

\textbf{Theorem 3.2 }\emph{The following equality holds good\ }%
\begin{equation*}
\begin{array}{c}
\left[ \tilde{\delta}_{\alpha },\tilde{\delta}_{\beta }\right] _{\left( \rho
,\eta \right) TE}=L_{\alpha \beta }^{\gamma }\circ \left( h\circ \pi \right)
\tilde{\delta}_{\gamma }+\left( \rho ,\eta ,h\right) \mathbb{R}_{\,\ \alpha
\beta }^{a}\overset{\cdot }{\tilde{\partial}}_{a},%
\end{array}%
\leqno(3.17)
\end{equation*}%
\emph{where }%
\begin{equation*}
\begin{array}{l}
\left( \rho ,\eta ,h\right) \mathbb{R}_{\,\ \alpha \beta }^{a}=\Gamma \left(
\tilde{\rho},Id_{E}\right) \left( \tilde{\delta}_{\beta }\right) \left(
\left( \rho ,\eta \right) \Gamma _{\alpha }^{a}\right) \vspace*{1mm} \\
\qquad -\Gamma \left( \tilde{\rho},Id_{E}\right) \left( \tilde{\delta}%
_{\alpha }\right) \left( \left( \rho ,\eta \right) \Gamma _{\beta
}^{a}\right) +\left( L_{\alpha \beta }^{\gamma }\circ h\circ \pi \right)
\left( \rho ,\eta \right) \Gamma _{\gamma }^{a},%
\end{array}%
\leqno(3.18)
\end{equation*}

\emph{Moreover, we have: }%
\begin{equation*}
\begin{array}{c}
\left[ \tilde{\delta}_{\alpha },\overset{\cdot }{\tilde{\partial}}_{b}\right]
_{\left( \rho ,\eta \right) TE}=\Gamma \left( \tilde{\rho},Id_{E}\right)
\left( \overset{\cdot }{\tilde{\partial}}_{b}\right) \left( \left( \rho
,\eta \right) \Gamma _{\alpha }^{a}\right) \overset{\cdot }{\tilde{\partial}}%
_{a},%
\end{array}%
\leqno(3.19)
\end{equation*}%
\emph{and}
\begin{equation*}
\begin{array}{c}
\Gamma \left( \tilde{\rho},Id_{E}\right) \left[ \tilde{\delta}_{\alpha },%
\tilde{\delta}_{\beta }\right] _{\left( \rho ,\eta \right) TE}=\left[ \Gamma
\left( \tilde{\rho},Id_{E}\right) \left( \tilde{\delta}_{\alpha }\right)
,\Gamma \left( \tilde{\rho},Id_{E}\right) \left( \tilde{\delta}_{\beta
}\right) \right] _{TE}.%
\end{array}%
\leqno(3.20)
\end{equation*}

Let $\left( d\tilde{z}^{\alpha },d\tilde{y}^{b}\right) $ be the natural dual
$\left( \rho ,\eta \right) $-base of natural $\left( \rho ,\eta \right) $%
-base $\left( \displaystyle\tilde{\partial}_{\alpha },\displaystyle\overset{%
\cdot }{\tilde{\partial}}_{a}\right) .$

This is determined by the equations
\begin{equation*}
\begin{array}{c}
\left\{
\begin{array}{cc}
\displaystyle\left\langle d\tilde{z}^{\alpha },\tilde{\partial}_{\beta
}\right\rangle =\delta _{\beta }^{\alpha }, & \displaystyle\left\langle d%
\tilde{z}^{\alpha },\overset{\cdot }{\tilde{\partial}}_{a}\right\rangle =0,%
\vspace*{2mm} \\
\displaystyle\left\langle d\tilde{y}^{a},\tilde{\partial}_{\beta
}\right\rangle =0, & \displaystyle\left\langle d\tilde{y}^{a},\overset{\cdot
}{\tilde{\partial}}_{b}\right\rangle =\delta _{b}^{a}.%
\end{array}%
\right.%
\end{array}%
\end{equation*}

We consider the problem of finding a base for the $\mathcal{F}\left(
E\right) $-module
\begin{equation*}
\left( \Gamma \left( \left( V\left( \rho ,\eta \right) TE\right) ^{\ast
},\left( \left( \rho ,\eta \right) \tau _{E}\right) ^{\ast },E\right)
,+,\cdot \right)
\end{equation*}%
of the type
\begin{equation*}
\begin{array}{c}
\delta \tilde{y}^{a}=\theta _{\alpha }^{a}d\tilde{z}^{\alpha }+\omega
_{b}^{a}d\tilde{y}^{b},~a\in \overline{1,n}%
\end{array}%
\end{equation*}%
which satisfies the following conditions:
\begin{equation*}
\begin{array}{c}
\left\langle \delta \tilde{y}^{a},\overset{\cdot }{\tilde{\partial}}%
_{a}\right\rangle =1\wedge \left\langle \delta \tilde{y}^{a},\tilde{\delta}%
_{\alpha }\right\rangle =0.%
\end{array}%
\leqno(3.21)
\end{equation*}

We obtain the sections
\begin{equation*}
\begin{array}{l}
\delta \tilde{y}^{a}=\left( \rho ,\eta \right) \Gamma _{\alpha }^{a}d\tilde{z%
}^{\alpha }+d\tilde{y}^{a},a\in \overline{1,n}.%
\end{array}%
\leqno(3.22)
\end{equation*}%
such that their changing rule is tensorial under a change of vector fiber
charts. The base $\left( d\tilde{z}^{\alpha },\delta \tilde{y}^{a}\right) $
will be called the \emph{adapted dual }$\left( \rho ,\eta \right) $\emph{%
-base.}

\section{The lift of a differentiable curve}

We consider the following diagram:
\begin{equation*}
\begin{array}{c}
\xymatrix{E\ar[d]_\pi&(F,[,]_{F,h},(\rho,\eta))\ar[d]^\nu\\ M\ar[r]^h&N}%
\end{array}%
\leqno(4.1)
\end{equation*}%
where $\left( E,\pi ,M\right) \in \left\vert \mathbf{B}^{\mathbf{v}%
}\right\vert $ and $\left( \left( F,\nu ,M\right) ,\left[ ,\right]
_{F,h},\left( \rho ,\eta \right) \right) \in \left\vert \mathbf{GLA}%
\right\vert .$

We admit that $\left( \rho ,\eta \right) \Gamma $ is a $\left( \rho ,\eta
\right) $-connection for the vector bundle $\left( E,\pi ,M\right) $ and $%
\begin{array}[b]{ccc}
I & ^{\underrightarrow{~c\ }} & M%
\end{array}%
$ is a differentiable curve. We know that
\begin{equation*}
\begin{array}{c}
\left( E_{|\func{Im}\left( \eta \circ h\circ c\right) },\pi _{|\func{Im}%
\left( \eta \circ h\circ c\right) },\func{Im}\left( \eta \circ h\circ
c\right) \right)%
\end{array}%
\end{equation*}%
is a vector subbundle of the vector bundle $\left( E,\pi ,M\right) .$

\textbf{Definition 4.1} If
\begin{equation*}
\begin{array}{ccc}
I & ^{\underrightarrow{\ \ \dot{c}\ \ }} & E_{|\func{Im}\left( \eta \circ
h\circ c\right) } \\
t & \longmapsto & y^{a}\left( t\right) s_{a}\left( \eta \circ h\circ c\left(
t\right) \right)%
\end{array}%
\leqno(4.2)
\end{equation*}%
is a differentiable curve such that there exists $g\in \mathbf{Man}\left(
E,F\right) $ such that the following conditions are satisfied:

\begin{itemize}
\item[1.] $\left( g,h\right) \in \mathbf{B}^{v}\left( \left( E,\pi ,M\right)
,\left( F,\nu ,N\right) \right) $ and

\item[2.] $\rho \circ g\circ \dot{c}\left( t\right) =\displaystyle\frac{%
d\left( \eta \circ h\circ c\right) ^{i}\left( t\right) }{dt}\frac{\partial }{%
\partial x^{i}}\left( \left( \eta \circ h\circ c\right) \left( t\right)
\right) ,$ for any $t\in I,$ \smallskip

then we will say that $\dot{c}$ \emph{\ is the }$\left( g,h\right) $\emph{%
-lift of the differentiable curve }$c.$
\end{itemize}

\emph{Remark 4.1 }The condition $2$ is equivalent with the following
affirmation:
\begin{equation*}
\begin{array}[b]{c}
\rho _{\alpha }^{i}\left( \eta \circ h\circ c\left( t\right) \right) \cdot
g_{a}^{\alpha }\left( h\circ c\left( t\right) \right) \cdot y^{a}\left(
t\right) =\frac{d\left( \eta \circ h\circ c\right) ^{i}\left( t\right) }{dt}%
,~i\in \overline{1,m}.%
\end{array}%
\leqno\left( 4.3\right)
\end{equation*}

\textbf{Definition 4.2} If $%
\begin{array}{ccc}
I & ^{\underrightarrow{\dot{c}}} & E_{|\func{Im}\left( \eta \circ h\circ
c\right) }%
\end{array}%
$ is a differentiable $\left( g,h\right) $-lift of the differentiable curve $%
c,$ then the section%
\begin{equation*}
\begin{array}{ccc}
\func{Im}\left( \eta \circ h\circ c\right) & ^{\underrightarrow{u\left( c,%
\dot{c}\right) }} & E_{|\func{Im}\left( \eta \circ h\circ c\right) }\vspace*{%
1mm} \\
\eta \circ h\circ c\left( t\right) & \longmapsto & \dot{c}\left( t\right)%
\end{array}%
\leqno(4.4)
\end{equation*}%
will be called the\emph{\ canonical section associated to the couple }$%
\left( c,\dot{c}\right) .$

\textbf{Definition 4.3 }If $\left( g,h\right) \in \mathbf{B}^{\mathbf{v}%
}\left( \left( E,\pi ,M\right) ,\left( F,\nu ,N\right) \right) $ has the
components
\begin{equation*}
\begin{array}{c}
g_{a}^{\alpha };a\in \overline{1,r},~\alpha \in \overline{1,p}%
\end{array}%
\end{equation*}%
such that for any vector local $\left( n+p\right) $-chart $\left(
V,t_{V}\right) $ of $\left( F,\nu ,N\right) $ there exists the real
functions
\begin{equation*}
\begin{array}{ccc}
V & ^{\underrightarrow{~\ \ \ \tilde{g}_{\alpha }^{a}~\ \ }} & \mathbb{R}%
\end{array}%
;~a\in \overline{1,r},~\alpha \in \overline{1,p}
\end{equation*}%
such that
\begin{equation*}
\begin{array}{c}
\tilde{g}_{\alpha }^{b}\left( \varkappa \right) \cdot g_{a}^{\alpha }\left(
\varkappa \right) =\delta _{a}^{b},%
\end{array}%
\leqno(4.5)
\end{equation*}%
for any $\varkappa \in V,$ then we will say that \emph{the }$\mathbf{B}^{%
\mathbf{v}}$\emph{-morphism }$\left( g,h\right) $\emph{\ is locally
invertible.}

\emph{Remark 4.2 }In particular, if $\left( Id_{TM},Id_{M},Id_{M}\right)
=\left( \rho ,\eta ,h\right) $ and the $\mathbf{B}^{\mathbf{v}}$ morphism $%
\left( g,Id_{M}\right) $ is locally invertible, then we have the
differentiable $\left( g,Id_{M}\right) $-lift
\begin{equation*}
\begin{array}{ccl}
I & ^{\underrightarrow{\ \ \dot{c}\ \ }} & TM \\
t & \longmapsto & \displaystyle\tilde{g}_{j}^{i}\left( c\left( t\right)
\right) \frac{dc^{j}\left( t\right) }{dt}\frac{\partial }{\partial x^{i}}%
\left( c\left( t\right) \right)%
\end{array}%
.\leqno(4.6)
\end{equation*}

Moreover, if $g=Id_{TM}$, then we obtain the usual lift of tangent vectors%
\begin{equation*}
\begin{array}{ccl}
I & ^{\underrightarrow{\ \ \dot{c}\ \ }} & TM\vspace*{1mm} \\
t & \longmapsto & \displaystyle\frac{dc^{i}\left( t\right) }{dt}\frac{%
\partial }{\partial x^{i}}\left( c\left( t\right) \right)%
\end{array}%
.\leqno(4.7)
\end{equation*}

\textbf{Definition 4.4 }If $%
\begin{array}{ccl}
I & ^{\underrightarrow{\ \ \dot{c}\ \ }} & E_{|\func{Im}\left( \eta \circ
h\circ c\right) }%
\end{array}%
$ is a differentiable $\left( g,h\right) $-lift of differentiable curve $c,$
such that its components functions $\left( y^{a},~a\in \overline{1,n}\right)
$ are solutions for the differentiable system of equations:%
\begin{equation*}
\begin{array}[b]{c}
\frac{du^{a}}{dt}+\left( \rho ,\eta \right) \Gamma _{\alpha }^{a}\circ
u\left( c,\dot{c}\right) \circ \left( \eta \circ h\circ c\right) \cdot
g_{b}^{\alpha }\circ h\circ c\cdot u^{b}=0,%
\end{array}%
\leqno(4.8)
\end{equation*}%
then we will say that \emph{the }$\left( g,h\right) $\emph{-lift }$\dot{c}$%
\emph{\ is parallel with respect to the }$\left( \rho ,\eta \right) $\emph{%
-connection }$\left( \rho ,\eta \right) \Gamma .$

\emph{Remark 4.3 }In particular, if $\left( \rho ,\eta ,h\right) =\left(
Id_{TM},Id_{M},Id_{M}\right) $ and the $\mathbf{B}^{\mathbf{v}}$ morphism $%
\left( g,Id_{M}\right) $ is locally invertible, then the differentiable $%
\left( g,Id_{TM}\right) $-lift
\begin{equation*}
\begin{array}{ccl}
I & ^{\underrightarrow{\ \ \dot{c}\ \ }} & TM\vspace*{1mm} \\
t & \longmapsto & \displaystyle\left( \tilde{g}_{j}^{i}\circ c\cdot \frac{%
dc^{j}}{dt}\right) \frac{\partial }{\partial x^{i}}\left( c\left( t\right)
\right) ,%
\end{array}%
\leqno(4.9)
\end{equation*}%
is parallel with respect to the connection $\Gamma $ if the component
functions
\begin{equation*}
\begin{array}[b]{c}
\left( \tilde{g}_{j}^{i}\circ c\cdot \frac{dc^{j}}{dt},~i\in \overline{1,n}%
\right)%
\end{array}%
\end{equation*}%
are solutions for the differentiable system of equations%
\begin{equation*}
\begin{array}[b]{c}
\frac{du^{i}}{dt}+\Gamma _{k}^{i}\circ u\left( c,\dot{c}\right) \circ c\cdot
g_{h}^{k}\circ c\cdot u^{h}=0,%
\end{array}%
\leqno(4.10)
\end{equation*}%
namely%
\begin{equation*}
\begin{array}{l}
\displaystyle\frac{d}{dt}\left( \tilde{g}_{j}^{i}\left( c\left( t\right)
\right) \cdot \frac{dc^{j}\left( t\right) }{dt}\right) \vspace*{1mm} \\
\qquad \displaystyle+\Gamma _{k}^{i}\left( \left( \tilde{g}_{j}^{i}\left(
c\left( t\right) \right) \cdot \frac{dc^{j}\left( t\right) }{dt}\right)
\cdot \frac{\partial }{\partial x^{i}}\left( c\left( t\right) \right)
\right) \cdot \frac{dc^{k}\left( t\right) }{dt}=0.%
\end{array}%
\leqno(4.10)^{\prime }
\end{equation*}

Moreover, if $g=Id_{TM}$, then the usual lift of tangent vectors $\left(
4.6\right) ^{\prime }$ is parallel with respect to the connection $\Gamma $
if the component functions $\left( \frac{dc^{j}}{dt},~j\in \overline{1,n}%
\right) $ are solutions for the differentiable system of equations
\begin{equation*}
\begin{array}[b]{c}
\frac{du^{i}}{dt}+\Gamma _{k}^{i}\circ u\left( c,\dot{c}\right) \circ c\cdot
u^{k}=0,%
\end{array}%
\leqno(4.11)
\end{equation*}%
namely%
\begin{equation*}
\begin{array}[b]{c}
\frac{d}{dt}\left( \frac{dc^{j}\left( t\right) }{dt}\right) +\Gamma
_{k}^{i}\left( \frac{dc^{j}\left( t\right) }{dt}\cdot \frac{\partial }{%
\partial x^{i}}\left( c\left( t\right) \right) \right) \cdot \frac{%
dc^{k}\left( t\right) }{dt}=0.%
\end{array}%
\leqno(4.11)^{\prime }
\end{equation*}

\section{Remarkable $\mathbf{Mod}$-endomorphisms}

In the following we consider the diagram:
\begin{equation*}
\begin{array}{c}
\xymatrix{E\ar[d]_\pi&\left( F,\left[ ,\right] _{F,h},\left( \rho ,\eta
\right) \right)\ar[d]^\nu\\ M\ar[r]^h&N}%
\end{array}%
\end{equation*}%
where $\left( E,\pi ,M\right) \in \left\vert \mathbf{B}^{\mathbf{v}%
}\right\vert $ and $\left( \left( F,\nu ,N\right) ,\left[ ,\right]
_{F,h},\left( \rho ,\eta \right) \right) $ is a generalized Lie algebroid.

\textbf{Definition 5.1 }For any $\mathbf{Mod}$-endomorphism $e$ of $\Gamma
\left( \left( \rho ,\eta \right) TE,\left( \rho ,\eta \right) \tau
_{E},E\right) $ we define the application of Nijenhuis \ type
\begin{equation*}
\!\!\Gamma \!((\rho ,\!\eta )TE,\!(\rho ,\eta )\tau _{E},E)^{2~\
\underrightarrow{~\ \ N_{e}~\ \ }}~\ \Gamma \!((\rho ,\eta )TE,\!(\rho ,\eta
)\tau _{E},\!E)\vspace*{1mm}
\end{equation*}%
defined by
\begin{equation*}
\begin{array}{c}
N_{e}\left( X,Y\right) =\left[ eX,eY\right] _{\rho TE}+e^{2}\left[ X,Y\right]
_{\rho TE}-e\left[ eX,Y\right] _{\rho TE}-e\left[ X,eY\right] _{\rho TE},%
\end{array}%
\end{equation*}%
for any $X,Y\in \Gamma \!((\rho ,\eta )TE,\!(\rho ,\eta )\tau _{E},\!E)%
\vspace*{1mm}.$

\subsection{Projectors}

\textbf{Definition 5.1.1 }Any $\mathbf{Mod}$-endomorphism $e$ of $\Gamma
\left( (\rho ,\eta )TE,\break (\rho ,\eta \ )\tau _{E},E\right) $ with the
property
\begin{equation*}
\begin{array}{l}
e^{2}=e%
\end{array}%
\leqno(5.1.1)
\end{equation*}%
will be called \emph{projector.}

\textbf{Example 5.1.1 }The $\mathbf{Mod}$-endomorphism
\begin{equation*}
\begin{array}{rcl}
\Gamma \left( \left( \rho ,\eta \right) TE,\left( \rho ,\eta \right) \tau
_{E},E\right) & ^{\underrightarrow{\ \ \mathcal{V}\ \ }} & \Gamma \left(
\left( \rho ,\eta \right) TE,\left( \rho ,\eta \right) \tau _{E},E\right) \\
Z^{\alpha }\tilde{\delta}_{\alpha }+Y^{a}\overset{\cdot }{\tilde{\partial}}%
_{a} & \longmapsto & Y^{a}\overset{\cdot }{\tilde{\partial}}_{a}%
\end{array}%
\end{equation*}%
is a projector which will be called \textit{the vertical projector.}

\emph{\ Remark 5.1.1 } We have $\mathcal{V}\left( \tilde{\delta}_{\alpha
}\right) =0$ and $\mathcal{V}\left( \overset{\cdot }{\tilde{\partial}}%
_{a}\right) =\overset{\cdot }{\tilde{\partial}}_{a}.$ Therefore, it follows%
\vspace*{-2mm}
\begin{equation*}
\mathcal{V}\left( \tilde{\partial}_{\alpha }\right) =\left( \rho ,\eta
\right) \Gamma _{\alpha }^{a}\overset{\cdot }{\tilde{\partial}}_{a}.
\end{equation*}

In addition, we obtain the equality%
\begin{equation*}
\begin{array}[b]{c}
\Gamma \left( \left( \rho ,\eta \right) \Gamma ,Id_{E}\right) \left(
Z^{\alpha }\tilde{\partial}_{\alpha }+Y^{a}\overset{\cdot }{\tilde{\partial}}%
_{a}\right) =\mathcal{V}\left( Z^{\alpha }\tilde{\partial}_{\alpha }+Y^{a}%
\overset{\cdot }{\tilde{\partial}}_{a}\right) ,%
\end{array}%
\leqno(5.1.2)
\end{equation*}%
for any $Z^{\alpha }\tilde{\partial}_{\alpha }+Y^{a}\overset{\cdot }{\tilde{%
\partial}}_{a}\in \Gamma \left( \left( \rho ,\eta \right) TE,\left( \rho
,\eta \right) \tau _{E},E\right) .$

\textbf{Theorem 5.1.1 }\emph{A }$(\rho ,\eta )$\emph{-connection for the
vector bundle} $(E,\pi ,M)$ \emph{is characterized by the existence of a} $%
\mathbf{Mod}$\emph{-endomorphism} $\mathcal{V}$ \emph{of }$\Gamma \left(
\left( \rho ,\eta \right) TE,\left( \rho ,\eta \right) \tau _{E},E\right) $
\emph{with the properties:}
\begin{equation*}
\begin{array}{c}
\mathcal{V}\left( \Gamma \left( \left( \rho ,\eta \right) TE,\left( \rho
,\eta \right) \tau _{E},E\right) \right) \subset \Gamma \left( V\left( \rho
,\eta \right) TE,\left( \rho ,\eta \right) \tau _{E},E\right) \vspace*{1mm}
\\
\mathcal{V}\left( X\right) =X\Longleftrightarrow X\in \Gamma \left( V\left(
\rho ,\eta \right) TE,\left( \rho ,\eta \right) \tau _{E},E\right)%
\end{array}%
\leqno(5.1.3)
\end{equation*}

\textbf{Example 5.1.2 }The $\mathbf{Mod}$-endomorphism
\begin{equation*}
\begin{array}{rcl}
\Gamma \left( \left( \rho ,\eta \right) TE,\left( \rho ,\eta \right) \tau
_{E},E\right) & ^{\underrightarrow{\ \ \mathcal{H}\ \ }} & \Gamma \left(
\left( \rho ,\eta \right) TE,\left( \rho ,\eta \right) \tau _{E},E\right) \\
Z^{\alpha }\tilde{\delta}_{\alpha }+Y^{a}\overset{\cdot }{\tilde{\partial}}%
_{a} & \longmapsto & Z^{\alpha }\tilde{\delta}_{\alpha }%
\end{array}%
\end{equation*}%
is a projector which will be called the \emph{horizontal projector.}

\emph{Remark 5.1.2} We have $\mathcal{H}\left( \tilde{\delta}_{\alpha
}\right) {=}\tilde{\delta}_{\alpha }$ and $\mathcal{H}\big(\overset{\cdot }{%
\tilde{\partial}}_{a}\big){=}0.$ Therefore, we obtain $\mathcal{H}\left(
\tilde{\partial}_{\alpha }\right) {=}\tilde{\delta}_{\alpha }.$

\textbf{Theorem 5.1.2 }\emph{A }$\left( \rho ,\eta \right) $\emph{%
-connection for the vector bundle} $\left( E,\pi ,M\right) $ \emph{\ is
characterized by the existence of a }$\mathbf{Mod}$\emph{-endomorphism }$%
\mathcal{H}$\emph{\ of }$\Gamma \left( \left( \rho ,\eta \right) TE,\left(
\rho ,\eta \right) \tau _{E},E\right) $ \emph{with the properties:}
\begin{equation*}
\begin{array}{c}
\mathcal{H}\left( \Gamma \left( \left( \rho ,\eta \right) TE,\left( \rho
,\eta \right) \tau _{E},E\right) \right) \subset \Gamma \left( H\left( \rho
,\eta \right) TE,\left( \rho ,\eta \right) \tau _{E},E\right) \vspace*{1mm}
\\
\mathcal{H}\left( X\right) =X\Longleftrightarrow X\in \Gamma \left( H\left(
\rho ,\eta \right) TE,\left( \rho ,\eta \right) \tau _{E},E\right) .%
\end{array}%
\leqno(5.1.4)
\end{equation*}

\textbf{Corollary 5.1.1} \emph{A }$\left( \rho ,\eta \right) $\emph{%
-connection for the vector bundle} $\left( E,\pi ,M\right) $ \emph{is
characterized by the existence of a }$\mathbf{Mod}$\emph{-endomorphism }$%
\mathcal{H}$\emph{\ of }$\Gamma \left( \left( \rho ,\eta \right) TE,\left(
\rho ,\eta \right) \tau _{E},E\right) $\emph{\ with the properties:}
\begin{equation*}
\begin{array}{c}
\mathcal{H}^{2}=\mathcal{H}\vspace*{1mm} \\
Ker\left( \mathcal{H}\right) =\left( \Gamma \left( V\left( \rho ,\eta
\right) TE,\left( \rho ,\eta \right) \tau _{E},E\right) ,+,\cdot \right) .%
\end{array}%
\leqno(5.1.5)
\end{equation*}

\emph{Remark 5.1.3 }For any $X\in \Gamma \left( \left( \rho ,\eta \right)
TE,\left( \rho ,\eta \right) \tau _{E},E\right) $ we obtain the unique
decomposition
\begin{equation*}
X=\mathcal{H}X+\mathcal{V}X.
\end{equation*}

\textbf{Proposition 5.1.1} \emph{After some calculations we obtain }%
\begin{equation*}
\begin{array}{c}
N_{\mathcal{V}}\left( X,Y\right) =\mathcal{V}\left[ \mathcal{H}X,\mathcal{H}Y%
\right] _{\left( \rho ,\eta \right) TE}=N_{\mathcal{H}}\left( X,Y\right) ,%
\end{array}%
\leqno(5.1.6)
\end{equation*}%
\textit{for any }$X,Y\in \Gamma \left( \left( \rho ,\eta \right) TE,\left(
\rho ,\eta \right) \tau _{E},E\right) .$

\textbf{Corollary 5.1.2} \emph{The horizontal interior differential system }%
\begin{equation*}
\left( H\left( \rho ,\eta \right) TE,\left( \rho ,\eta \right) \tau
_{E},E\right)
\end{equation*}%
\emph{is involutive if and only if }$N_{\mathcal{V}}=0$\emph{\ or }$N_{%
\mathcal{H}}=0.$

\subsection{The almost product structure}

\textbf{Definition 5.2.1 }Any $\mathbf{Mod}$-endomorphism $e$ of $\Gamma
\left( (\rho ,\eta )TE,\break (\rho ,\eta \ )\tau _{E},E\right) $ with the
property%
\begin{equation*}
\begin{array}{c}
e^{2}=Id%
\end{array}%
\leqno(5.2.1)
\end{equation*}%
will be called the \emph{almost product structure}.

\textbf{Example 5.2.1 }The $\mathbf{Mod}$-endomorphism
\begin{equation*}
\begin{array}{rcl}
\Gamma \left( \left( \rho ,\eta \right) TE,\left( \rho ,\eta \right) \tau
_{E},E\right) & ^{\underrightarrow{\ \ \mathcal{P}\ \ }} & \Gamma \left(
\left( \rho ,\eta \right) TE,\left( \rho ,\eta \right) \tau _{E},E\right) \\
Z^{\alpha }\tilde{\delta}_{\alpha }+Y^{a}\overset{\cdot }{\tilde{\partial}}%
_{a} & \longmapsto & Z^{\alpha }\tilde{\delta}_{\alpha }-Y^{a}\overset{\cdot
}{\tilde{\partial}}_{a}%
\end{array}%
\end{equation*}%
is an almost product structure.

\emph{\ Remark 5.2.1 }The previous almost product structure has the
properties:
\begin{equation*}
\begin{array}{l}
\mathcal{P}=\left( 2\mathcal{H}-Id\right) ; \\
\mathcal{P}=\left( Id-2\mathcal{V}\right) ; \\
\mathcal{P}=\left( \mathcal{H}-\mathcal{V}\right) .%
\end{array}%
\leqno(5.2.2)
\end{equation*}

\emph{Remark 5.2.2 } We obtain that $\mathcal{P}\left( \tilde{\delta}%
_{\alpha }\right) =\tilde{\delta}_{\alpha }$ and $\mathcal{P}\left( \overset{%
\cdot }{\tilde{\partial}}_{a}\right) =-\overset{\cdot }{\tilde{\partial}}%
_{a}.$ Therefore, it follows \vspace*{-2mm}
\begin{equation*}
\mathcal{P}\left( \tilde{\partial}_{\alpha }\right) =\tilde{\delta}_{\alpha
}-\rho \Gamma _{\alpha }^{a}\overset{\cdot }{\tilde{\partial}}_{a}.
\end{equation*}

\textbf{Theorem 5.2.1 }\emph{A }$\left( \rho ,\eta \right) $\emph{%
-connection for the vector bundle }$\left( E,\pi ,M\right) $\emph{\ is
characterized by the existence of a }$\mathbf{Mod}$\emph{-endomorphism }$%
\mathcal{P}$\emph{\ of }$\Gamma \left( \left( \rho ,\eta \right) TE,\left(
\rho ,\eta \right) \tau _{E},E\right) $ \emph{with the following property:}
\begin{equation*}
\begin{array}{c}
\mathcal{P}\left( X\right) =-X\Longleftrightarrow X\in \Gamma \left( V\left(
\rho ,\eta \right) TE,\left( \rho ,\eta \right) \tau _{E},E\right) .%
\end{array}%
\leqno(5.2.3)
\end{equation*}

\textbf{Proposition 5.2.1} \emph{After some calculations, we obtain }%
\begin{equation*}
N_{\mathcal{P}}\left( X,Y\right) =4\mathcal{V}\left[ \mathcal{H}X,\mathcal{H}%
Y\right] ,
\end{equation*}%
\emph{for any }$X,Y\in \Gamma \left( \left( \rho ,\eta \right) TE,\left(
\rho ,\eta \right) \tau _{E},E\right) .$

\textbf{Corollary 5.2.1} \emph{The horizontal interior differential system }$%
\left( H\left( \rho ,\eta \right) TE,\left( \rho ,\eta \right) \tau
_{E},E\right) $ \emph{is involutive if and only if }$N_{\mathcal{P}}=0.$

\subsection{\noindent The almost tangent structure}

\textbf{Definition 5.3.1 }Any $\mathbf{Mod}$-endomorphism $e$ of $\left(
\Gamma \!((\rho ,\eta )TE,\break (\rho ,\eta )\tau _{E},E\right) $ with the
property
\begin{equation*}
\begin{array}{c}
e^{2}=0%
\end{array}%
\leqno(5.3.1)
\end{equation*}%
will be called the \emph{almost tangent structure.}

\textbf{Example 5.3.1 }If $\left( E,\pi ,M\right) =\left( F,\nu ,N\right) $,
$g\in \mathbf{Man}\left( E,E\right) $ such that $\left( g,h\right) $ is a
locally invertible $\mathbf{B}^{\mathbf{v}}$\textit{-}morphism, then the $%
\mathbf{Mod}$-endomorphism
\begin{equation*}
\begin{array}{rcl}
\Gamma \left( \left( \rho ,\eta \right) TE,\left( \rho ,\eta \right) \tau
_{E},E\right) & ^{\underrightarrow{\mathcal{J}_{\left( g,h\right) }}} &
\Gamma \left( \left( \rho ,\eta \right) TE,\left( \rho ,\eta \right) \tau
_{E},E\right) \\
Z^{a}\tilde{\delta}_{a}+Y^{b}\overset{\cdot }{\tilde{\partial}}_{b} &
\longmapsto & \left( \tilde{g}_{a}^{b}\circ h\circ \pi \right) Z^{a}\overset{%
\cdot }{\tilde{\partial}}_{b}%
\end{array}%
\end{equation*}%
is an almost tangent structure which will be called the \emph{almost tangent
structure associated to the }$\mathbf{B}^{\mathbf{v}}$\emph{-morphism }$%
\left( g,h\right) $. (See: \emph{Definition 4.3}\textbf{)}

\emph{Remark 5.3.1 }We obtain that
\begin{equation*}
\mbox{$\mathcal{J}_{\left( g,h\right)
}\left( \tilde{\delta}_{a}\right) =\mathcal{J}_{\left( g,h\right) }\left(
\tilde{\partial}_{a}\right) =\left( \tilde{g}_{a}^{b}\circ h\circ \pi
\right) \overset{\cdot }{\tilde{\partial}}_{b}$ and $\mathcal{J}_{\left(
g,h\right) }\left( \overset{\cdot }{\tilde{\partial}}_{b}\right) =0.$}
\end{equation*}%
and we have the following properties:
\begin{equation*}
\begin{array}{rcl}
\mathcal{J}_{\left( g,h\right) }\circ \mathcal{P} & = & \mathcal{J}_{\left(
g,h\right) };\vspace*{1mm} \\
\mathcal{P}\circ \mathcal{J}_{\left( g,h\right) } & = & -\mathcal{J}_{\left(
g,h\right) };\vspace*{1mm} \\
\mathcal{J}_{\left( g,h\right) }\circ \mathcal{H} & = & \mathcal{J}_{\left(
g,h\right) };\vspace*{1mm} \\
\mathcal{H}\circ \mathcal{J}_{\left( g,h\right) } & = & 0;\vspace*{1mm} \\
\mathcal{J}_{\left( g,h\right) }\circ \mathcal{V} & = & 0;\vspace*{1mm} \\
\mathcal{V}\circ \mathcal{J}_{\left( g,h\right) } & = & \mathcal{J}_{\left(
g,h\right) };\vspace*{1mm} \\
N_{\mathcal{J}_{\left( g,h\right) }} & = & 0.%
\end{array}%
\leqno(5.3.2)
\end{equation*}

\section{Tensor $d$-fields. Distinguished linear $\left( \protect\rho ,%
\protect\eta \right) $-connections}

We consider the following diagram:
\begin{equation*}
\begin{array}{c}
\xymatrix{E\ar[d]_\pi&\left( F,\left[ ,\right] _{F,h},\left( \rho ,\eta
\right) \right)\ar[d]^\nu\\ M\ar[r]^h&N}%
\end{array}%
\end{equation*}%
where $\left( E,\pi ,M\right) \in \left\vert \mathbf{B}^{\mathbf{v}%
}\right\vert $ and $\left( \left( F,\nu ,N\right) ,\left[ ,\right]
_{F,h},\left( \rho ,\eta \right) \right) $ is a generalized Lie algebroid.

Let
\begin{equation*}
\left( \mathcal{T}~_{q,s}^{p,r}\left( \left( \rho ,\eta \right) TE,\left(
\rho ,\eta \right) \tau _{E},E\right) ,+,\cdot \right)
\end{equation*}%
be the $\mathcal{F}\left( E\right) $-module of tensor fields by $\left(
_{q,s}^{p,r}\right) $-type from the generalized tangent bundle
\begin{equation*}
\left( H\left( \rho ,\eta \right) TE,\left( \rho ,\eta \right) \tau
_{E},E\right) \oplus \left( V\left( \rho ,\eta \right) TE,\left( \rho ,\eta
\right) \tau _{E},E\right) .
\end{equation*}

An arbitrarily tensor field $T$ is written as
\begin{equation*}
\begin{array}{c}
T=T_{\beta _{1}...\beta _{q}b_{1}...b_{s}}^{\alpha _{1}...\alpha
_{p}a_{1}...a_{r}}\tilde{\delta}_{\alpha _{1}}\otimes ...\otimes \tilde{%
\delta}_{\alpha _{p}}\otimes d\tilde{z}^{\beta _{1}}\otimes ...\otimes d%
\tilde{z}^{\beta _{q}}\otimes \\
\overset{\cdot }{\tilde{\partial}}_{a_{1}}\otimes ...\otimes \overset{\cdot }%
{\tilde{\partial}}_{a_{r}}\otimes \delta \tilde{y}^{b_{1}}\otimes ...\otimes
\delta \tilde{y}^{b_{s}}.%
\end{array}%
\end{equation*}

Let
\begin{equation*}
\left( \mathcal{T}~\left( \left( \rho ,\eta \right) TE,\left( \rho ,\eta
\right) \tau _{E},E\right) ,+,\cdot ,\otimes \right)
\end{equation*}%
be the tensor fields algebra of generalized tangent bundle $\left( \left(
\rho ,\eta \right) TE,\left( \rho ,\eta \right) \tau _{E},E\right) $.

If $T_{1} \in \mathcal{T}_{q_{1},s_{1}}^{p_{1},r_{1}}(( \rho ,\eta ) TE,(
\rho ,\eta ) \tau _{E},E) $ and $T_{2}{\in} \mathcal{T}%
_{q_{2},s_{2}}^{p_{2},r_{2}}( ( \rho ,\eta ) TE,( \rho ,\eta ) \tau _{E},E) $%
, then the components of product tensor field $T_{1}\otimes T_{2}$ are the
products of local components of $T_{1}$ and $T_{2}.$

Therefore, we obtain $T_{1}\otimes T_{2}\in \mathcal{T}%
_{q_{1}+q_{2},s_{1}+s_{2}}^{p_{1}+p_{2},r_{1}+r_{2}}\left( \left( \rho ,\eta
\right) TE,\left( \rho ,\eta \right) \tau _{E},E\right) .$\smallskip

Let $\mathcal{DT}( ( \rho ,\eta ) TE,( \rho ,\eta ) \tau _{E},E) $ be the
family of tensor fields
\begin{equation*}
T\in \mathcal{T}( ( \rho ,\eta ) TE,( \rho ,\eta ) \tau _{E},E)
\end{equation*}
for which there exists
\begin{equation*}
\mbox{$T_{1}\in \mathcal{T}_{q,0}^{p,0}(
( \rho ,\eta ) TE,( \rho ,\eta ) \tau _{E},E) $
and $T_{2}\in \mathcal{T}_{0,s}^{0,r}( ( \rho ,\eta )
TE,( \rho ,\eta ) \tau _{E},E) $}
\end{equation*}
such that $T=T_{1}+T_{2}.$

The $\mathcal{F}\left( E\right) $-module $\left( \mathcal{DT}\left( \left(
\rho ,\eta \right) TE,\left( \rho ,\eta \right) \tau _{E},E\right) ,+,\cdot
\right) $ will be called the \emph{module of distinguished tensor fields} or
the \emph{module of tensor }$d$-\emph{fields.}

\emph{\ Remark 5.1 }The elements of
\begin{equation*}
\Gamma \left( \left( \rho ,\eta \right) TE,\left( \rho ,\eta \right) \tau
_{E},E\right)
\end{equation*}%
respectively
\begin{equation*}
\Gamma (((\rho ,\eta )TE)^{\ast },\break ((\rho ,\eta )\tau _{E})^{\ast },E)
\end{equation*}%
are tensor $d$-fields.

\textbf{Definition 6.1 }Let $\left( E,\pi ,M\right) $ be a vector bundle
endowed with a $\left( \rho ,\eta \right) $-connection $\left( \rho ,\eta
\right) \Gamma $ and let
\begin{equation*}
\begin{array}{l}
\left( X,T\right) ^{\ \underrightarrow{\left( \rho ,\eta \right) D}\,}%
\vspace*{1mm}\left( \rho ,\eta \right) D_{X}T%
\end{array}%
\leqno(6.1)
\end{equation*}%
be a covariant $\left( \rho ,\eta \right) $-derivative for the tensor
algebra of the generalized tangent bundle
\begin{equation*}
\left( \left( \rho ,\eta \right) TE,\left( \rho ,\eta \right) \tau
_{E},E\right)
\end{equation*}%
which preserves the horizontal and vertical \emph{IDS} by parallelism.

The real local functions
\begin{equation*}
\left( \left( \rho ,\eta \right) H_{\beta \gamma }^{\alpha },\left( \rho
,\eta \right) H_{b\gamma }^{a},\left( \rho ,\eta \right) V_{\beta c}^{\alpha
},\left( \rho ,\eta \right) V_{bc}^{a}\right)
\end{equation*}%
defined by the following equalities:
\begin{equation*}
\begin{array}{ll}
\left( \rho ,\eta \right) D_{\tilde{\delta}_{\gamma }}\tilde{\delta}_{\beta
}=\left( \rho ,\eta \right) H_{\beta \gamma }^{\alpha }\tilde{\delta}%
_{\alpha }, & \left( \rho ,\eta \right) D_{\tilde{\delta}_{\gamma }}\overset{%
\cdot }{\tilde{\partial}}_{b}=\left( \rho ,\eta \right) H_{b\gamma }^{a}%
\overset{\cdot }{\tilde{\partial}}_{a} \\
\left( \rho ,\eta \right) D_{\overset{\cdot }{\tilde{\partial}}_{c}}\tilde{%
\delta}_{\beta }=\left( \rho ,\eta \right) V_{\beta c}^{\alpha }\tilde{\delta%
}_{\alpha }, & \left( \rho ,\eta \right) D_{\overset{\cdot }{\tilde{\partial}%
}_{c}}\overset{\cdot }{\tilde{\partial}}_{b}=\left( \rho ,\eta \right)
V_{bc}^{a}\overset{\cdot }{\tilde{\partial}}_{a}%
\end{array}%
\leqno(6.2)
\end{equation*}%
are the components of a linear $\left( \rho ,\eta \right) $-connection
\begin{equation*}
\left( \left( \rho ,\eta \right) H,\left( \rho ,\eta \right) V\right)
\end{equation*}%
for the generalized tangent bundle $\left( \left( \rho ,\eta \right)
TE,\left( \rho ,\eta \right) \tau _{E},E\right) $ which will be called the
\emph{distinguished linear }$\left( \rho ,\eta \right) $\emph{-connection.}

If $h=Id_{M},$ then the distinguished linear $\left( Id_{TM},Id_{M}\right) $%
-connection will be called the \emph{distinguished linear connection.}

The components of a distinguished linear connection $\left( H,V\right) $
will be denoted
\begin{equation*}
\left( H_{jk}^{i},H_{bk}^{a},V_{jc}^{i},V_{bc}^{a}\right) .
\end{equation*}

\textbf{Theorem 6.1 }\emph{If }$((\rho ,\eta )H,(\rho ,\eta )V)$ \emph{is a
distinguished linear} $(\rho ,\eta )$-\emph{connection for the generalized
tangent bundle }$\left( \left( \rho ,\eta \right) TE,\left( \rho ,\eta
\right) \tau _{E},E\right) $\emph{, then its components satisfy the change
relations: }

\begin{equation*}
\begin{array}{ll}
\left( \rho ,\eta \right) H_{\beta
{\acute{}}%
\gamma
{\acute{}}%
}^{\alpha
{\acute{}}%
}\!\! & =\Lambda _{\alpha }^{\alpha
{\acute{}}%
}\circ h\circ \pi \cdot \left[ \Gamma \left( \tilde{\rho},Id_{E}\right)
\left( \tilde{\delta}_{\gamma }\right) \left( \Lambda _{\beta
{\acute{}}%
}^{\alpha }\circ h\circ \pi \right) +\right. \vspace*{1mm} \\
& +\left. \left( \rho ,\eta \right) H_{\beta \gamma }^{\alpha }\cdot \Lambda
_{\beta
{\acute{}}%
}^{\beta }\circ h\circ \pi \right] \cdot \Lambda _{\gamma
{\acute{}}%
}^{\gamma }\circ h\circ \pi ,\vspace*{2mm} \\
\left( \rho ,\eta \right) H_{b%
{\acute{}}%
\gamma
{\acute{}}%
}^{a%
{\acute{}}%
}\!\! & =M_{a}^{a%
{\acute{}}%
}\circ \pi \cdot \left[ \Gamma \left( \tilde{\rho},Id_{E}\right) \left(
\tilde{\delta}_{\gamma }\right) \left( M_{b%
{\acute{}}%
}^{a}\circ \pi \right) +\right. \vspace*{1mm} \\
& \left. +\left( \rho ,\eta \right) H_{b\gamma }^{a}\cdot M_{b%
{\acute{}}%
}^{b}\circ \pi \right] \cdot \Lambda _{\gamma
{\acute{}}%
}^{\gamma }\circ h\circ \pi ,\vspace*{2mm} \\
\left( \rho ,\eta \right) V_{\beta
{\acute{}}%
c%
{\acute{}}%
}^{\alpha
{\acute{}}%
}\!\! & =\Lambda _{\alpha
{\acute{}}%
}^{\alpha }\circ h\circ \pi \cdot \left( \rho ,\eta \right) V_{\beta
c}^{\alpha }\cdot \Lambda _{\beta
{\acute{}}%
}^{\beta }\circ h\circ \pi \cdot M_{c%
{\acute{}}%
}^{c}\circ \pi ,\vspace*{2mm} \\
\left( \rho ,\eta \right) V_{b%
{\acute{}}%
c%
{\acute{}}%
}^{a%
{\acute{}}%
}\!\! & =M_{a}^{a%
{\acute{}}%
}\circ \pi \cdot \left( \rho ,\eta \right) V_{bc}^{a}\cdot M_{b%
{\acute{}}%
}^{b}\circ \pi \cdot M_{c%
{\acute{}}%
}^{c}\circ \pi .%
\end{array}%
\leqno(6.3)
\end{equation*}

\emph{The components of a distinguished linear connection $\left( H,V\right)
$ verify the change relations:}
\begin{equation*}
\begin{array}{cl}
H_{j%
{\acute{}}%
k%
{\acute{}}%
}^{i%
{\acute{}}%
} & =\frac{\partial x^{i%
{\acute{}}%
}}{\partial x^{i}}\circ \pi \cdot \left[ \displaystyle\frac{\delta }{\delta
x^{k}}\left( \displaystyle\frac{\partial x^{i}}{\partial x^{j%
{\acute{}}%
}}\circ \pi \right) +H_{jk}^{i}\cdot \displaystyle\frac{\partial x^{j}}{%
\partial x^{j%
{\acute{}}%
}}\circ \pi \right] \cdot \displaystyle\frac{\partial x^{k}}{\partial x^{k%
{\acute{}}%
}}\circ \pi ,\vspace*{2mm} \\
H_{b%
{\acute{}}%
k%
{\acute{}}%
}^{a%
{\acute{}}%
} & =M_{a}^{a%
{\acute{}}%
}\circ \pi \cdot \left[ \displaystyle\frac{\delta }{\delta x^{k}}\left( M_{b%
{\acute{}}%
}^{a}\circ \pi \right) +H_{bk}^{a}\cdot M_{b%
{\acute{}}%
}^{b}\circ \pi \right] \cdot \displaystyle\frac{\partial x^{k}}{\partial x^{k%
{\acute{}}%
}}\circ \pi ,\vspace*{2mm} \\
V_{j%
{\acute{}}%
c%
{\acute{}}%
}^{i%
{\acute{}}%
} & =\displaystyle\frac{\partial x^{i%
{\acute{}}%
}}{\partial x^{i}}\circ \pi \cdot V_{jc}^{i}\displaystyle\frac{\partial x^{j}%
}{\partial x^{j%
{\acute{}}%
}}\circ \pi \cdot M_{c%
{\acute{}}%
}^{c}\circ \pi ,\vspace*{3mm} \\
V_{b%
{\acute{}}%
c%
{\acute{}}%
}^{a%
{\acute{}}%
} & =M_{a}^{a%
{\acute{}}%
}\circ \pi \cdot V_{bc}^{a}\cdot M_{b%
{\acute{}}%
}^{b}\circ \pi \cdot M_{c%
{\acute{}}%
}^{c}\circ \pi .%
\end{array}%
\leqno(6.3)^{\prime }
\end{equation*}

\textbf{Example 6.1} If $\left( E,\pi ,M\right) $ is a vector bundle endowed
with the $\left( \rho ,\eta \right) $-connection $\left( \rho ,\eta \right)
\Gamma $, then the local real functions%
\begin{equation*}
\begin{array}[b]{c}
\left( \frac{\partial \left( \rho ,\eta \right) \Gamma _{\gamma }^{a}}{%
\partial y^{b}},\frac{\partial \left( \rho ,\eta \right) \Gamma _{\gamma
}^{a}}{\partial y^{b}},0,0\right)
\end{array}%
\leqno(6.4)
\end{equation*}%
are the components of a distinguished linear $\left( \rho ,\eta \right) $%
\textit{-}connection for $\left( \left( \rho ,\eta \right) TE,\left( \rho
,\eta \right) \tau _{E},E\right) ,$ which will by called the \emph{Berwald
linear }$\left( \rho ,\eta \right) $\emph{-connection.}

The Berwald linear $(Id_{TM},Id_{M})$-connection will be called the \emph{%
Ber\-wald linear connection.}

\textbf{Theorem 6.2} \emph{If the generalized tangent bundle} $\!(\!(\rho
,\!\eta )T\!E,\!(\rho ,\!\eta )\tau _{E},\!E\!)$ \emph{is endowed with a
distinguished linear} $\!(\rho ,\!\eta )$\emph{-connection} $((\rho ,\eta
)H,(\rho ,\eta )V)$, \emph{then, for any}
\begin{equation*}
\begin{array}[b]{c}
X=\tilde{Z}^{\alpha }\tilde{\delta}_{\alpha }+Y^{a}\overset{\cdot }{\tilde{%
\partial}}_{a}\in \Gamma (\!(\rho ,\eta )TE,\!(\rho ,\!\eta )\tau _{E},\!E)%
\end{array}%
\end{equation*}%
\emph{and for any}
\begin{equation*}
T\in \mathcal{T}_{qs}^{pr}\!(\!(\rho ,\eta )TE,\!(\rho ,\eta )\tau _{E},\!E),
\end{equation*}%
\emph{we obtain the formula:}
\begin{equation*}
\begin{array}{l}
\left( \rho ,\eta \right) D_{X}\left( T_{\beta _{1}...\beta
_{q}b_{1}...b_{s}}^{\alpha _{1}...\alpha _{p}a_{1}...a_{r}}\tilde{\delta}%
_{\alpha _{1}}\otimes ...\otimes \tilde{\delta}_{\alpha _{p}}\otimes d\tilde{%
z}^{\beta _{1}}\otimes ...\otimes \right. \vspace*{1mm} \\
\hspace*{9mm}\left. \otimes d\tilde{z}^{\beta _{q}}\otimes \overset{\cdot }{%
\tilde{\partial}}_{a_{1}}\otimes ...\otimes \overset{\cdot }{\tilde{\partial}%
}_{a_{r}}\otimes \delta \tilde{y}^{b_{1}}\otimes ...\otimes \delta \tilde{y}%
^{b_{s}}\right) =\vspace*{1mm} \\
\hspace*{9mm}=\tilde{Z}^{\gamma }T_{\beta _{1}...\beta _{q}b_{1}...b_{s}\mid
\gamma }^{\alpha _{1}...\alpha _{p}a_{1}...a_{r}}\tilde{\delta}_{\alpha
_{1}}\otimes ...\otimes \tilde{\delta}_{\alpha _{p}}\otimes d\tilde{z}%
^{\beta _{1}}\otimes ...\otimes d\tilde{z}^{\beta _{q}}\otimes \overset{%
\cdot }{\tilde{\partial}}_{a_{1}}\otimes ...\otimes \vspace*{1mm} \\
\hspace*{9mm}\otimes \overset{\cdot }{\tilde{\partial}}_{a_{r}}\otimes
\delta \tilde{y}^{b_{1}}\otimes ...\otimes \delta \tilde{y}%
^{b_{s}}+Y^{c}T_{\beta _{1}...\beta _{q}b_{1}...b_{s}}^{\alpha _{1}...\alpha
_{p}a_{1}...a_{r}}\mid _{c}\tilde{\delta}_{\alpha _{1}}\otimes ...\otimes
\vspace*{1mm} \\
\hspace*{9mm}\otimes \tilde{\delta}_{\alpha _{p}}\otimes d\tilde{z}^{\beta
_{1}}\otimes ...\otimes d\tilde{z}^{\beta _{q}}\otimes \overset{\cdot }{%
\tilde{\partial}}_{a_{1}}\otimes ...\otimes \overset{\cdot }{\tilde{\partial}%
}_{a_{r}}\otimes \delta \tilde{y}^{b_{1}}\otimes ...\otimes \delta \tilde{y}%
^{b_{s}},%
\end{array}%
\end{equation*}%
\emph{where }%
\begin{equation*}
\begin{array}{l}
T_{\beta _{1}...\beta _{q}b_{1}...b_{s}\mid \gamma }^{\alpha _{1}...\alpha
_{p}a_{1}...a_{r}}=\vspace*{2mm}\Gamma \left( \tilde{\rho},Id_{E}\right)
\left( \tilde{\delta}_{\gamma }\right) T_{\beta _{1}...\beta
_{q}b_{1}...b_{s}}^{\alpha _{1}...\alpha _{p}a_{1}...a_{r}} \\
\hspace*{8mm}+\left( \rho ,\eta \right) H_{\alpha \gamma }^{\alpha
_{1}}T_{\beta _{1}...\beta _{q}b_{1}...b_{s}}^{\alpha \alpha _{2}...\alpha
_{p}a_{1}...a_{r}}+...+\vspace*{2mm}\left( \rho ,\eta \right) H_{\alpha
\gamma }^{\alpha _{p}}T_{\beta _{1}...\beta _{q}b_{1}...b_{s}}^{\alpha
_{1}...\alpha _{p-1}\alpha a_{1}...a_{r}} \\
\hspace*{8mm}-\left( \rho ,\eta \right) H_{\beta _{1}\gamma }^{\beta
}T_{\beta \beta _{2}...\beta _{q}b_{1}...b_{s}}^{\alpha _{1}...\alpha
_{p}a_{1}...a_{r}}-...-\vspace*{2mm}\left( \rho ,\eta \right) H_{\beta
_{q}\gamma }^{\beta }T_{\beta _{1}...\beta _{q-1}\beta
b_{1}...b_{s}}^{\alpha _{1}...\alpha _{p}a_{1}...a_{r}} \\
\hspace*{8mm}+\left( \rho ,\eta \right) H_{a\gamma }^{a_{1}}T_{\beta
_{1}...\beta _{q}b_{1}...b_{s}}^{\alpha _{1}...\alpha
_{p}aa_{2}...a_{r}}+...+\vspace*{2mm}\left( \rho ,\eta \right) H_{a\gamma
}^{a_{r}}T_{\beta _{1}...\beta _{q}b_{1}...b_{s}}^{\alpha _{1}...\alpha
_{p}a_{1}...a_{r-1}a} \\
\hspace*{8mm}-\left( \rho ,\eta \right) H_{b_{1}\gamma }^{b}T_{\beta
_{1}...\beta _{q}bb_{2}...b_{s}}^{\alpha _{1}...\alpha _{p}a_{1}...a_{r}}-%
\vspace*{2mm}...-\left( \rho ,\eta \right) H_{b_{s}\gamma }^{b}T_{\beta
_{1}...\beta _{q}b_{1}...b_{s-1}b}^{\alpha _{1}...\alpha _{p}a_{1}...a_{r}}%
\end{array}%
\end{equation*}%
\emph{and }%
\begin{equation*}
\begin{array}{l}
T_{\beta _{1}...\beta _{q}b_{1}...b_{s}}^{\alpha _{1}...\alpha
_{p}a_{1}...a_{r}}\mid _{c}=\Gamma \left( \tilde{\rho},Id_{E}\right) \left(
\overset{\cdot }{\tilde{\partial}}_{c}\right) T_{\beta _{1}...\beta
_{q}b_{1}...b_{s}}^{\alpha _{1}...\alpha _{p}a_{1}...a_{r}}+\vspace*{2mm} \\
\hspace*{8mm}+\left( \rho ,\eta \right) V_{\alpha c}^{\alpha _{1}}T_{\beta
_{1}...\beta _{q}b_{1}...b_{s}}^{\alpha \alpha _{2}...\alpha
_{p}a_{1}...a_{r}}+...+\left( \rho ,\eta \right) V_{\alpha c}^{\alpha
_{p}}T_{\beta _{1}...\beta _{q}b_{1}...b_{s}}^{\alpha _{1}...\alpha
_{p-1}\alpha a_{1}...a_{r}}-\vspace*{2mm} \\
\hspace*{8mm}-\left( \rho ,\eta \right) V_{\beta _{1}c}^{\beta }T_{\beta
\beta _{2}...\beta _{q}b_{1}...b_{s}}^{\alpha _{1}...\alpha
_{p}a_{1}...a_{r}}-...-\left( \rho ,\eta \right) V_{\beta _{q}c}^{\beta
}T_{\beta _{1}...\beta _{q-1}\beta b_{1}...b_{s}}^{\alpha _{1}...\alpha
_{p}a_{1}...a_{r}}+\vspace*{2mm} \\
\hspace*{8mm}+\left( \rho ,\eta \right) V_{ac}^{a_{1}}T_{\beta _{1}...\beta
_{q}b_{1}...b_{s}}^{\alpha _{1}...\alpha _{p}aa_{2}...a_{r}}+...+\left( \rho
,\eta \right) V_{ac}^{a_{r}}T_{\beta _{1}...\beta _{q}b_{1}...b_{s}}^{\alpha
_{1}...\alpha _{p}a_{1}...a_{r-1}a}-\vspace*{2mm} \\
\hspace*{8mm}-\left( \rho ,\eta \right) V_{b_{1}c}^{b}T_{\beta _{1}...\beta
_{q}bb_{2}...b_{s}-}^{\alpha _{1}...\alpha _{p}a_{1}...a_{r}}...-\left( \rho
,\eta \right) V_{b_{s}c}^{b}T_{\beta _{1}...\beta
_{q}b_{1}...b_{s-1}b}^{\alpha _{1}...\alpha _{p}a_{1}...a_{r}}.%
\end{array}%
\end{equation*}

\textbf{Definition 6.2 }We assume that $\left( E,\pi ,M\right) =\left( F,\nu
,N\right) .$

If $\left( \rho ,\eta \right) \Gamma $ is a $\left( \rho ,\eta \right) $%
-connection for the vector bundle $\left( E,\pi ,M\right) $ and
\begin{equation*}
\left( \left( \rho ,\eta \right) H_{bc}^{a},\left( \rho ,\eta \right) \tilde{%
H}_{bc}^{a},\left( \rho ,\eta \right) V_{bc}^{a},\left( \rho ,\eta \right)
\tilde{V}_{bc}^{a}\right)
\end{equation*}%
are the components of a distinguished linear $\left( \rho ,\eta \right) $%
\textit{-}connection for the generalized tangent bundle $\left( \left( \rho
,\eta \right) TE,\left( \rho ,\eta \right) \tau _{E},E\right) $ such that
\begin{equation*}
\left( \rho ,\eta \right) H_{bc}^{a}=\left( \rho ,\eta \right) \tilde{H}%
_{bc}^{a}\mbox{ and }\left( \rho ,\eta \right) V_{bc}^{a}=\left( \rho ,\eta
\right) \tilde{V}_{bc}^{a},
\end{equation*}%
then we will say that \emph{the generalized tangent bundle }$\!(\! ( \rho
,\!\eta ) TE, ( \rho ,\!\eta ) \tau _{E},\!E ) $ \emph{is endowed with a
normal distinguished linear }$\left( \rho ,\eta \right) $\emph{-connection
on components }$\left( \left( \rho ,\eta \right) H_{bc}^{a},\left( \rho
,\eta \right) V_{bc}^{a}\right) $.

The components of a normal distinguished linear $\left(
Id_{TM},Id_{M}\right) $-con\-nec\-tion $\left( H,V\right) $ will be denoted $%
\left( H_{jk}^{i},V_{jk}^{i}\right) $.

\bigskip

\section{Mechanical systems}

We consider the following diagram:
\begin{equation*}
\begin{array}{c}
\xymatrix{E\ar[d]_\pi&\left( E,\left[ ,\right] _{E,h},\left( \rho ,\eta
\right) \right)\ar[d]^\pi \\ M\ar[r]^h&M}%
\end{array}%
\leqno(7.1)
\end{equation*}%
where $\left( \left( E,\pi ,M\right) ,\left[ ,\right] _{E,h},\left( \rho
,\eta \right) \right) $ is a generalized Lie algebroid.

\textbf{Definition 7.1 }A triple
\begin{equation*}
\begin{array}{c}
\left( \left( E,\pi ,M\right) ,F_{e},\left( \rho ,\eta \right) \Gamma
\right) ,%
\end{array}%
\leqno(7.2)
\end{equation*}%
where
\begin{equation*}
\begin{array}[t]{l}
F_{e}=F^{a}\frac{\partial }{\partial \tilde{y}^{a}}\in \Gamma \left( V\left(
\rho ,\eta \right) TE,\left( \rho ,\eta \right) \tau _{E},E\right)
\end{array}%
\leqno(7.3)
\end{equation*}%
is an external force and $\left( \rho ,\eta \right) \Gamma $ is a $\left(
\rho ,\eta \right) $-connection for the vector bundle $\left( E,\pi
,M\right) $, will be called \emph{mechanical }$\left( \rho ,\eta \right) $%
\emph{-system.}

\textbf{Definition 7.2 }A smooth \emph{Lagrange fundamental function} on the
vector bundle\break $\left( E,\pi ,M\right) $ is a mapping $E~\ ^{%
\underrightarrow{\ \ L\ \ }}~\ \mathbb{R}$ which satisfies the following
conditions:\medskip

1. $L\circ u\in C^{\infty }\left( M\right) $, for any $u\in \Gamma \left(
E,\pi ,M\right) \setminus \left\{ 0\right\} $;\smallskip

2. $L\circ 0\in C^{0}\left( M\right) $, where $0$ means the null section of $%
\left( E,\pi ,M\right) .$\medskip

Let $L$ be a Lagrangian defined on the total space of the vector bundle $%
\left( E,\pi ,M\right) .$

If $\left( U,s_{U}\right) $ is a local vector $\left( m+r\right) $-chart for
$\left( E,\pi ,M\right) $, then we obtain the following real functions
defined on $\pi ^{-1}\left( U\right) $:%
\begin{equation*}
\begin{array}{cc}
L_{i}\overset{put}{=}\displaystyle\frac{\partial L}{\partial x^{i}}\overset{%
put}{=}\frac{\partial }{\partial x^{i}}\left( L\right)  & L_{ib}\overset{put}%
{=}\displaystyle\frac{\partial ^{2}L}{\partial x^{i}\partial y^{b}}\vspace*{%
2mm}\overset{put}{=}\frac{\partial }{\partial x^{i}}\left( \frac{\partial }{%
\partial y^{b}}\left( L\right) \right)  \\
L_{a}\overset{put}{=}\displaystyle\frac{\partial L}{\partial y^{a}}\overset{%
put}{=}\frac{\partial }{\partial y^{a}}\left( L\right)  & L_{ab}\overset{put}%
{=}\displaystyle\frac{\partial ^{2}L}{\partial y^{a}\partial y^{b}}\overset{%
put}{=}\frac{\partial }{\partial y^{a}}\left( \frac{\partial }{\partial y^{b}%
}\left( L\right) \right) .%
\end{array}%
\leqno(7.4)
\end{equation*}

\textbf{Definition 7.3 }If for any vector local $m+r$-chart $\left(
U,s_{U}\right) $ of $\left( E,\pi ,M\right) ,$ we have:
\begin{equation*}
\begin{array}{c}
rank\left\Vert L_{ab}\left( u_{x}\right) \right\Vert =r,%
\end{array}%
\leqno(7.5)
\end{equation*}%
for any $u_{x}\in \pi ^{-1}\left( U\right) \backslash \left\{ 0_{x}\right\} $%
, then we will say that \emph{the Lagrangian }$L$\emph{\ is regular.}

\textbf{Proposition 7.1} \emph{If the Lagrangian }$L$\emph{\ is regular,
then for any vector local }$m+r$\emph{-chart }$\left( U,s_{U}\right) $\emph{%
\ of }$\left( E,\pi ,M\right) ,$\emph{\ we obtain the real functions }$%
\tilde{L}^{ab}$\emph{\ locally defined by}%
\begin{equation*}
\begin{array}{ccc}
\pi ^{-1}\left( U\right)  & ^{\underrightarrow{\ \ \tilde{L}^{ab}\ \ }} &
\mathbb{R} \\
u_{x} & \longmapsto  & \tilde{L}^{ab}\left( u_{x}\right)
\end{array}%
,\leqno(7.6)
\end{equation*}%
\emph{where }$\left\Vert \tilde{L}^{ab}\left( u_{x}\right) \right\Vert
=\left\Vert L_{ab}\left( u_{x}\right) \right\Vert ^{-1}$\emph{, for any }$%
u_{x}\in \pi ^{-1}\left( U\right) \backslash \left\{ 0_{x}\right\} .$

\textbf{Definition 7.4 }A smooth \emph{Finsler fundamental function} on the
vector bundle $\left( E,\pi ,M\right) $ is a mapping $%
\begin{array}[b]{ccc}
E & ^{\underrightarrow{\ F\ }} & \mathbb{R}_{+}%
\end{array}%
$ which satisfies the following conditions:\medskip

1. $F\circ u\in C^{\infty }\left( M\right) $, for any $u\in \Gamma \left(
E,\pi ,M\right) \setminus \left\{ 0\right\} $;\smallskip

2. $F\circ 0\in C^{0}\left( M\right) $, where $0$ means the null section of $%
\left( E,\pi ,M\right) $;\smallskip

3. $F$ is positively $1$-homogenous on the fibres of vector bundle $\left(
E,\pi ,M\right) ;$\smallskip

4. For any vector local $m+r$-chart $\left( U,s_{U}\right) $ of $\left(
E,\pi ,M\right) ,$ the hessian:%
\begin{equation*}
\left\Vert F_{~ab}^{2}\left( u_{x}\right) \right\Vert \leqno(7.7)
\end{equation*}%
is positively define for any $u_{x}\in \pi ^{-1}\left( U\right) \backslash
\left\{ 0_{x}\right\} $.

\textbf{Definition 7.5 }If $L$ (respectively $F$) is a smooth Lagrange
(respectively Finsler function), then the triple
\begin{equation*}
\begin{array}[t]{l}
\left( \left( E,\pi ,M\right) ,F_{e},L\right)%
\end{array}%
\mbox{\ \ \ (
respectively
$
\left( \left( E,\pi ,M\right),F_{e},F\right))
$}
\end{equation*}%
where $F_{e}=F^{a}\displaystyle\frac{\partial }{\partial \tilde{y}^{a}}\in
\Gamma \left( V\left( \rho ,\eta \right) TE,\left( \rho ,\eta \right) \tau
_{E},E\right) $ is an external force, is called \emph{Lagrange mechanical }$%
\left( \rho ,\eta \right) $\emph{-system }\ and \emph{Finsler mechanical }$%
\left( \rho ,\eta \right) $\emph{-system, respectively}.

Any Lagrange mechanical $\left( Id_{TM},Id_{M}\right) $-system and any
Finsler mechanical\break $\left( Id_{TM},Id_{M}\right) $-system will be
called \emph{Lagrange mechanical system }and \emph{Finsler mechanical system}%
, respectively.

\section{ $(\protect\rho ,\protect\eta )$-semisprays and $(\protect\rho ,%
\protect\eta )$-sprays for mechanical $(\protect\rho ,\protect\eta )$-systems%
}

Let $(\left( E,\pi ,M\right) ,F_{e},(\rho ,\eta )\Gamma )$ be an arbitrary
mechanical $\left( \rho ,\eta \right) $-system.

\textbf{Definition 8.1 }The\textit{\ }vertical section $\mathbb{C}\mathbf{=}%
y^{a}\overset{\cdot }{\tilde{\partial}}_{a}$will be called the\textit{\ }%
\emph{Liouville section.}

A section $S\in \Gamma \left( \left( \rho ,\eta \right) TE,\left( \rho ,\eta
\right) \tau _{E},E\right) $\ will be called $\left( \rho ,\eta \right) $%
\emph{-semispray}\ if there exists an almost tangent structure $e$ such that
$e\left( S\right) =\mathbb{C}.$

Let $g\in \mathbf{Man}\left( E,E\right) $ be such that $\left( g,h\right) $
is a locally invertible $\mathbf{B}^{\mathbf{v}}$-morphism of $\left( E,\pi
,M\right) $\ source and\ target.

\textbf{Theorem 8.1 }\emph{The section }%
\begin{equation*}
\begin{array}[t]{l}
S=\left( g_{b}^{a}\circ h\circ \pi \right) y^{b}\tilde{\partial}_{a}-2\left(
G^{a}-\frac{1}{4}F^{a}\right) \overset{\cdot }{\tilde{\partial}}_{a}%
\end{array}%
\leqno(8.1)
\end{equation*}%
\emph{is a }$\left( \rho ,\eta \right) $\emph{-semispray such that the real
local functions }$G^{a},\ a\in \overline{1,n},$\emph{\ satisfy the following
conditions}%
\begin{equation*}
\begin{array}{cl}
\left( \rho ,\eta \right) \Gamma _{c}^{a} & =\left( \tilde{g}_{c}^{b}\circ
h\circ \pi \right) \frac{\partial \left( G^{a}-\frac{1}{4}F^{a}\right) }{%
\partial y^{b}} \\
& -\frac{1}{2}\left( g_{e}^{d}\circ h\circ \pi \right) y^{e}L_{dc}^{f}\left(
\tilde{g}_{f}^{a}\circ h\circ \pi \right)  \\
& +\frac{1}{2}\left( \rho _{c}^{j}\circ h\circ \pi \right) \frac{\partial
\left( g_{e}^{b}\circ h\circ \pi \right) }{\partial x^{j}}y^{e}\left( \tilde{%
g}_{b}^{a}\circ h\circ \pi \right)  \\
& -\frac{1}{2}\left( g_{e}^{b}\circ h\circ \pi \right) y^{e}\left( \rho
_{b}^{i}\circ h\circ \pi \right) \frac{\partial \left( \tilde{g}%
_{c}^{a}\circ h\circ \pi \right) }{\partial x^{i}}%
\end{array}%
\leqno(8.2)
\end{equation*}%
\emph{\ }

\emph{In addition, we remark that the local real functions}%
\begin{equation*}
\begin{array}{cl}
\left( \rho ,\eta \right) \mathring{\Gamma}_{c}^{a} & =\left( \tilde{g}%
_{c}^{b}\circ h\circ \pi \right) \frac{\partial G^{a}}{\partial y^{b}} \\
& -\frac{1}{2}\left( g_{e}^{d}\circ h\circ \pi \right) y^{e}L_{dc}^{b}\left(
\tilde{g}_{b}^{a}\circ h\circ \pi \right)  \\
& +\frac{1}{2}\left( \rho _{c}^{j}\circ h\circ \pi \right) \frac{\partial
\left( g_{e}^{b}\circ h\circ \pi \right) }{\partial x^{j}}y^{e}\left( \tilde{%
g}_{b}^{a}\circ h\circ \pi \right)  \\
& -\frac{1}{2}\left( g_{e}^{b}\circ h\circ \pi \right) y^{e}\left( \rho
_{b}^{i}\circ h\circ \pi \right) \frac{\partial \left( \tilde{g}%
_{c}^{a}\circ h\circ \pi \right) }{\partial x^{i}}%
\end{array}%
\leqno(8.3)
\end{equation*}%
\emph{are the components of a }$\left( \rho ,\eta \right) $\emph{-connection
}$\left( \rho ,\eta \right) \mathring{\Gamma}$\emph{\ for the vector bundle }%
$\left( E,\pi ,M\right) .$

The $\left( \rho ,\eta \right) $-semispray $S$\ will be called \emph{the\
canonical }$\left( \rho ,\eta \right) $\emph{-semispray associated to
mechanical }$\left( \rho ,\eta \right) $\emph{-system }$\left( \left( E,\pi
,M\right) ,F_{e},\left( \rho ,\eta \right) \Gamma \right) $\emph{\ and from
locally invertible }$\mathbf{B}^{\mathbf{v}}$\emph{-morphism }$\left(
g,h\right) .$

\emph{Proof.} We consider the $\mathbf{Mod}$-endomorphism%
\begin{equation*}
\begin{array}{rcl}
\Gamma \left( \left( \rho ,\eta \right) TE,\left( \rho ,\eta \right) \tau
_{E},E\right) & ^{\underrightarrow{\ \ \mathbb{P}\ \ }} & \Gamma \left(
\left( \rho ,\eta \right) TE,\left( \rho ,\eta \right) \tau _{E},E\right)
\vspace*{1,5mm} \\
X & \longmapsto & \mathcal{J}_{\left( g,h\right) }\left[ S,X\right] _{\left(
\rho ,\eta \right) TE}-\left[ S,\mathcal{J}_{\left( g,h\right) }X\right]
_{\left( \rho ,\eta \right) TE}.%
\end{array}%
\end{equation*}

Let $X=Z^{a}\tilde{\partial}_{a}+Y^{a}\overset{\cdot }{\tilde{\partial}}_{a}$
be an arbitrary section. Since
\begin{equation*}
\begin{array}{cl}
\left[ S,X\right] _{\left( \rho ,\eta \right) TE} & =\displaystyle\left[
\left( g_{e}^{a}\circ h\circ \pi \cdot y^{e}\right) \tilde{\partial}%
_{a},Z^{b}\tilde{\partial}_{b}\right] _{\left( \rho ,\eta \right) TE}+\left[
\left( g_{e}^{a}\circ h\circ \pi \cdot y^{e}\right) \tilde{\partial}%
_{a},Y^{b}\overset{\cdot }{\tilde{\partial}}_{b}\right] _{\left( \rho ,\eta
\right) TE}\vspace*{2mm} \\
& \displaystyle-\left[ 2\left( G^{a}-\frac{1}{4}F^{a}\right) \overset{\cdot }%
{\tilde{\partial}}_{a},Z^{b}\tilde{\partial}_{b}\right] _{\left( \rho ,\eta
\right) TE}-\left[ 2\left( G^{a}-\frac{1}{4}F^{a}\right) \overset{\cdot }{%
\tilde{\partial}}_{a},Y^{b}\overset{\cdot }{\tilde{\partial}}_{b}\right]
_{\left( \rho ,\eta \right) TE}%
\end{array}%
\end{equation*}%
and
\begin{equation*}
\begin{array}{cl}
\left[ \left( g_{e}^{a}\circ h\circ \pi \cdot y^{e}\right) \tilde{\partial}%
_{a},Z^{b}\tilde{\partial}_{b}\right] _{\left( \rho ,\eta \right) TE} & =%
\displaystyle\left( g_{e}^{a}\circ h\circ \pi \right) y^{e}\left( \rho
_{a}^{i}\circ h\circ \pi \right) \frac{\partial Z^{c}}{\partial x^{i}}\tilde{%
\partial}_{c}\vspace*{2mm} \\
& -\displaystyle Z^{b}\left( \rho _{b}^{j}\circ h\circ \pi \right) \frac{%
\partial \left( g_{e}^{c}\circ h\circ \pi \right) }{\partial x^{j}}y^{e}%
\tilde{\partial}_{c}\vspace*{2mm} \\
& \displaystyle+\left( g_{e}^{a}\circ h\circ \pi \right) y^{e}Z^{b}L_{ab}^{c}%
\tilde{\partial}_{c},%
\end{array}%
\end{equation*}%
\begin{equation*}
\begin{array}{cl}
\left[ \left( g_{e}^{a}\circ h\circ \pi \cdot y^{e}\right) \tilde{\partial}%
_{a},Y^{b}\overset{\cdot }{\tilde{\partial}}_{b}\right] _{\left( \rho ,\eta
\right) TE} & =\displaystyle\left( g_{e}^{a}\circ h\circ \pi \right)
y^{e}\left( \rho _{a}^{i}\circ h\circ \pi \right) \frac{\partial Y^{c}}{%
\partial x^{i}}\overset{\cdot }{\tilde{\partial}}_{c}\vspace*{2mm} \\
& \displaystyle-Y^{b}g_{b}^{c}\tilde{\partial}_{c},%
\end{array}%
\end{equation*}%
\begin{equation*}
\begin{array}{cl}
\displaystyle\left[ 2\left( G^{a}-\frac{1}{4}F^{a}\right) \overset{\cdot }{%
\tilde{\partial}}_{a},Z^{b}\tilde{\partial}_{b}\right] _{\left( \rho ,\eta
\right) TE} & \displaystyle=2\left( G^{a}-\frac{1}{4}F^{a}\right) \frac{%
\partial Z^{c}}{\partial y^{a}}\tilde{\partial}_{c}\vspace*{2mm} \\
& \displaystyle-2Z^{b}\rho _{b}^{j}\circ h\circ \pi \frac{\partial \left(
G^{c}-\frac{1}{4}F^{c}\right) }{\partial x^{j}}\overset{\cdot }{\tilde{%
\partial}}_{c},%
\end{array}%
\end{equation*}%
\begin{equation*}
\begin{array}{cl}
\left[ 2\left( G^{a}-\frac{1}{4}F^{a}\right) \overset{\cdot }{\tilde{\partial%
}}_{a},Y^{b}\overset{\cdot }{\tilde{\partial}}_{b}\right] _{\left( \rho
,\eta \right) TE}=2\left( G^{a}-\frac{1}{4}F^{a}\right) \frac{\partial Y^{c}%
}{\partial y^{a}}\overset{\cdot }{\tilde{\partial}}_{c}-2Y^{b}\displaystyle%
\frac{\partial \left( G^{c}-\displaystyle\frac{1}{4}F^{c}\right) }{\partial
y^{b}}\overset{\cdot }{\tilde{\partial}}_{c}, &
\end{array}%
\end{equation*}%
it results that
\begin{equation*}
\begin{array}{cl}
\mathcal{J}_{\left( g,h\right) }\left[ S,X\right] _{\left( \rho ,\eta
\right) TE} & \displaystyle=\left( g_{e}^{a}\circ h\circ \pi \right)
y^{e}\left( \rho _{a}^{i}\circ h\circ \pi \right) \frac{\partial Z^{c}}{%
\partial x^{i}}\left( \tilde{g}_{c}^{d}\circ h\circ \pi \right) \overset{%
\cdot }{\tilde{\partial}}_{d}\vspace*{2mm} \\
& \displaystyle-Z^{b}\left( \rho _{b}^{j}\circ h\circ \pi \right) \frac{%
\partial \left( g_{e}^{c}\circ h\circ \pi \right) }{\partial x^{j}}%
y^{e}\left( \tilde{g}_{c}^{d}\circ h\circ \pi \right) \overset{\cdot }{%
\tilde{\partial}}_{d}\vspace*{2mm} \\
& \displaystyle+\left( g_{e}^{a}\circ h\circ \pi \right)
y^{e}Z^{b}L_{ab}^{c}\left( \tilde{g}_{c}^{d}\circ h\circ \pi \right) \overset%
{\cdot }{\tilde{\partial}}_{d}\vspace*{2mm}\displaystyle-Y^{d}\overset{\cdot
}{\tilde{\partial}}_{d}\vspace*{2mm} \\
& \displaystyle-2\left( G^{a}-\frac{1}{4}F^{a}\right) \frac{\partial Z^{c}}{%
\partial y^{a}}\left( \tilde{g}_{c}^{d}\circ h\circ \pi \right) \overset{%
\cdot }{\tilde{\partial}}_{d}.%
\end{array}%
\leqno\left( P_{1}\right)
\end{equation*}

Since
\begin{equation*}
\begin{array}{cl}
\left[ S,\mathcal{J}_{\left( g,h\right) }X\right] _{\left( \rho ,\eta
\right) TE} & \displaystyle=\left[ \left( g_{e}^{a}\circ h\circ \pi \right)
y^{e}\tilde{\partial}_{a},Z^{b}\left( \tilde{g}_{b}^{c}\circ h\circ \pi
\right) \overset{\cdot }{\tilde{\partial}}_{c}\right] _{\left( \rho ,\eta
\right) TE}\vspace*{2mm} \\
& \displaystyle-\left[ 2\left( G^{a}-\frac{1}{4}F^{a}\right) \overset{\cdot }%
{\tilde{\partial}}_{a},Z^{b}\left( \tilde{g}_{b}^{c}\circ h\circ \pi \right)
\overset{\cdot }{\tilde{\partial}}_{c}\right] _{\left( \rho ,\eta \right) TE}%
\end{array}%
\end{equation*}%
and
\begin{equation*}
\begin{array}{cl}
\left[ \left( g_{e}^{a}\circ h\circ \pi \right) y^{e}\tilde{\partial}%
_{a},Z^{b}\left( \tilde{g}_{b}^{c}\circ h\circ \pi \right) \overset{\cdot }{%
\tilde{\partial}}_{c}\right] _{\left( \rho ,\eta \right) TE} & \displaystyle%
=-Z^{d}\tilde{\partial}_{d}+\left( g_{e}^{a}\circ h\circ \pi \right)
y^{e}\left( \rho _{a}^{i}\circ h\circ \pi \right) \frac{\partial Z^{b}}{%
\partial x^{i}}\left( \tilde{g}_{b}^{d}\circ h\circ \pi \right) \overset{%
\cdot }{\tilde{\partial}}_{d}\vspace*{2mm} \\
& \displaystyle-\left( g_{e}^{a}\circ h\circ \pi \right) y^{e}\left( \rho
_{a}^{i}\circ h\circ \pi \right) Z^{b}\frac{\partial \left( \tilde{g}%
_{b}^{d}\circ h\circ \pi \right) }{\partial x^{i}}\overset{\cdot }{\tilde{%
\partial}}_{d},%
\end{array}%
\end{equation*}%
\begin{equation*}
\begin{array}{cl}
\displaystyle\left[ 2\left( G^{a}-\frac{1}{4}F^{a}\right) \overset{\cdot }{%
\tilde{\partial}}_{a},Z^{b}\left( \tilde{g}_{b}^{c}\circ h\circ \pi \right)
\overset{\cdot }{\tilde{\partial}}_{c}\right] _{\left( \rho ,\eta \right) TE}
& \displaystyle=2\left( G^{a}-\frac{1}{4}F^{a}\right) \frac{\partial Z^{b}}{%
\partial y^{a}}\left( \tilde{g}_{b}^{d}\circ h\circ \pi \right) \overset{%
\cdot }{\tilde{\partial}}_{d}\vspace*{2mm} \\
& \displaystyle-Z^{b}\left( \tilde{g}_{b}^{c}\circ h\circ \pi \right) \frac{%
\partial 2\left( G^{d}-\frac{1}{4}F^{d}\right) }{\partial y^{c}}\overset{%
\cdot }{\tilde{\partial}}_{d}%
\end{array}%
\end{equation*}%
it results that%
\begin{equation*}
\begin{array}{cl}
\left[ S,\mathcal{J}_{\left( g,h\right) }X\right] _{\left( \rho ,\eta
\right) TE} & \displaystyle=-Z^{d}\tilde{\partial}_{d}+\left( g_{e}^{a}\circ
h\circ \pi \right) y^{e}\left( \rho _{a}^{i}\circ h\circ \pi \right) \frac{%
\partial Z^{b}}{\partial x^{i}}\left( \tilde{g}_{b}^{d}\circ h\circ \pi
\right) \overset{\cdot }{\tilde{\partial}}_{d} \\
& \displaystyle-\left( g_{e}^{a}\circ h\circ \pi \right) y^{e}\left( \rho
_{a}^{i}\circ h\circ \pi \right) Z^{b}\frac{\partial \left( \tilde{g}%
_{b}^{d}\circ h\circ \pi \right) }{\partial x^{i}}\overset{\cdot }{\tilde{%
\partial}}_{d} \\
& \displaystyle-2\left( G^{a}-\frac{1}{4}F^{a}\right) \frac{\partial Z^{b}}{%
\partial y^{a}}\left( \tilde{g}_{b}^{d}\circ h\circ \pi \right) \overset{%
\cdot }{\tilde{\partial}}_{d}\vspace*{2mm} \\
& \displaystyle+Z^{b}\left( \tilde{g}_{b}^{c}\circ h\circ \pi \right) \frac{%
\partial 2\left( G^{d}-\frac{1}{4}F^{d}\right) }{\partial y^{c}}\overset{%
\cdot }{\tilde{\partial}}_{d}.%
\end{array}%
\leqno\left( P_{2}\right)
\end{equation*}

Using equalities $\left( P_{1}\right) $ and $\left( P_{2}\right) $, we
obtain:%
\begin{equation*}
\begin{array}[b]{cl}
\mathbb{P}\left( Z^{a}\tilde{\partial}_{a}+Y^{a}\overset{\cdot }{\tilde{%
\partial}}_{a}\right) & =Z^{a}\tilde{\partial}_{a}-Y^{d}\overset{\cdot }{%
\tilde{\partial}}_{d}+\left( g_{e}^{a}\circ h\circ \pi \right)
y^{e}Z^{b}L_{ab}^{c}\left( \tilde{g}_{c}^{d}\circ h\circ \pi \right) \overset%
{\cdot }{\tilde{\partial}}_{d} \\
& -Z^{b}\left( \rho _{b}^{j}\circ h\circ \pi \right) \frac{\partial \left(
g_{e}^{c}\circ h\circ \pi \right) }{\partial x^{j}}y^{e}\left( \tilde{g}%
_{c}^{d}\circ h\circ \pi \right) \overset{\cdot }{\tilde{\partial}}_{d} \\
& +\left( g_{e}^{a}\circ h\circ \pi \right) y^{e}\left( \rho _{a}^{i}\circ
h\circ \pi \right) Z^{b}\frac{\partial \left( \tilde{g}_{b}^{d}\circ h\circ
\pi \right) }{\partial x^{i}}\overset{\cdot }{\tilde{\partial}}_{d} \\
& -Z^{b}\left( \tilde{g}_{b}^{c}\circ h\circ \pi \right) \frac{\partial
2\left( G^{d}-\frac{1}{4}F^{d}\right) }{\partial y^{c}}\overset{\cdot }{%
\tilde{\partial}}_{d}%
\end{array}%
\end{equation*}

After some calculations, it results that $\mathbb{P}$ is an almost product
structure.

Using the equalities $\left( 5.1.2\right) $ and $\left( 5.2.2\right) $ it
results that
\begin{equation*}
\mathbb{P}\left( Z^{a}\tilde{\partial}_{a}+Y^{a}\overset{\cdot }{\tilde{%
\partial}}_{a}\right) =\left( Id-2\left( \rho ,\eta \right) \Gamma \right)
\left( Z^{a}\tilde{\partial}_{a}+Y^{a}\overset{\cdot }{\tilde{\partial}}%
_{a}\right) ,
\end{equation*}%
for any $Z^{a}\tilde{\partial}_{a}+Y^{a}\overset{\cdot }{\tilde{\partial}}%
_{a}\in \Gamma \left( \left( \rho ,\eta \right) TE,\left( \rho ,\eta \right)
\tau _{E},E\right) $ and we obtain%
\begin{equation*}
\begin{array}[b]{cl}
\left( \rho ,\eta \right) \Gamma \left( Z^{a}\tilde{\partial}_{a}+Y^{a}%
\overset{\cdot }{\tilde{\partial}}_{a}\right) & =Y^{d}\overset{\cdot }{%
\tilde{\partial}}_{d}-\frac{1}{2}\left( g_{e}^{a}\circ h\circ \pi \right)
y^{e}Z^{b}L_{ab}^{c}\left( \tilde{g}_{c}^{d}\circ h\circ \pi \right) \overset%
{\cdot }{\tilde{\partial}}_{d} \\
& +\frac{1}{2}Z^{b}\left( \rho _{b}^{j}\circ h\circ \pi \right) \frac{%
\partial \left( g_{e}^{c}\circ h\circ \pi \right) }{\partial x^{j}}%
y^{e}\left( \tilde{g}_{c}^{d}\circ h\circ \pi \right) \overset{\cdot }{%
\tilde{\partial}}_{d} \\
& -\frac{1}{2}\left( g_{e}^{a}\circ h\circ \pi \right) y^{e}\left( \rho
_{a}^{i}\circ h\circ \pi \right) Z^{b}\frac{\partial \left( \tilde{g}%
_{b}^{d}\circ h\circ \pi \right) }{\partial x^{i}}\overset{\cdot }{\tilde{%
\partial}}_{d} \\
& +Z^{b}\left( \tilde{g}_{b}^{c}\circ h\circ \pi \right) \frac{\partial
\left( G^{d}-\frac{1}{4}F^{d}\right) }{\partial y^{c}}\overset{\cdot }{%
\tilde{\partial}}_{d}.%
\end{array}%
\end{equation*}

Since
\begin{equation*}
\begin{array}{c}
\left( \rho ,\eta \right) \Gamma \left( Z^{a}\tilde{\partial}_{a}+Y^{a}%
\overset{\cdot }{\tilde{\partial}}_{a}\right) =\left( Y^{d}+\left( \rho
,\eta \right) \Gamma _{b}^{d}Z^{b}\right) \overset{\cdot }{\tilde{\partial}}%
_{d}%
\end{array}%
\end{equation*}%
it results the relations $\left( 8.3\right) $. In addition, since
\begin{equation*}
\left( \rho ,\eta \right) \mathring{\Gamma}_{c}^{a}=\left( \rho ,\eta
\right) \Gamma _{c}^{a}+\frac{1}{4}\tilde{g}_{c}^{d}\circ h\circ \pi \frac{%
\partial F^{a}}{\partial y^{d}}
\end{equation*}%
and
\begin{equation*}
\begin{array}{ll}
\left( \rho ,\eta \right) \mathring{\Gamma}_{c%
{\acute{}}%
}^{a%
{\acute{}}%
} & \displaystyle=\left( \rho ,\eta \right) \Gamma _{c%
{\acute{}}%
}^{a%
{\acute{}}%
}+\frac{1}{2}\tilde{g}_{c%
{\acute{}}%
}^{b%
{\acute{}}%
}\circ h\circ \pi \displaystyle\frac{\partial F^{a%
{\acute{}}%
}}{\partial y^{b%
{\acute{}}%
}}\vspace*{2mm} \\
& \displaystyle=M_{a}^{a%
{\acute{}}%
}\circ \pi \left( \rho _{c}^{i}\circ h\circ \pi \cdot \frac{\partial M_{b%
{\acute{}}%
}^{a}}{\partial x^{i}}y^{b%
{\acute{}}%
}+\left( \rho ,\eta \right) \Gamma _{c}^{a}\right) M_{c%
{\acute{}}%
}^{c}\circ h\circ \pi \vspace*{2mm} \\
& \displaystyle+M_{a}^{a%
{\acute{}}%
}\circ \pi \left( \frac{1}{4}\tilde{g}_{c}^{b}\circ h\circ \pi \cdot \frac{%
\partial F^{a}}{\partial y^{b}}\right) M_{c%
{\acute{}}%
}^{c}\circ h\circ \pi \vspace*{2mm} \\
& \displaystyle=M_{a}^{a%
{\acute{}}%
}{\circ }\pi \left( \rho _{c}^{i}{\circ }h{\circ }\pi \cdot \frac{\partial
M_{b%
{\acute{}}%
}^{a}}{\partial x^{i}}y^{b%
{\acute{}}%
}+\left( \left( \rho ,\eta \right) \Gamma _{c}^{a}+\frac{1}{4}\tilde{g}%
_{c}^{b}{\circ h\circ }\pi \cdot \frac{\partial F^{a}}{\partial y^{b}}%
\right) \right) M_{c%
{\acute{}}%
}^{c}{\circ }h{\circ }\pi \vspace*{2mm} \\
& \displaystyle=M_{a}^{a%
{\acute{}}%
}{\circ }\pi \left( \rho _{c}^{i}{\circ }h{\circ }\pi \cdot \frac{\partial
M_{b%
{\acute{}}%
}^{a}}{\partial x^{i}}y^{b%
{\acute{}}%
}+\left( \rho ,\eta \right) \mathring{\Gamma}_{c}^{a}\right) M_{c%
{\acute{}}%
}^{c}{\circ }h{\circ }\pi%
\end{array}%
\end{equation*}%
it results the conclusion of the theorem. \hfill \emph{q.e.d.}

\emph{Remark 8.1 }If $\left( \rho ,\eta \right) =\left(
Id_{TM},Id_{M}\right) $, $\left( g,h\right) =\left( Id_{E},Id_{M}\right) $,\
and $F_{e}\neq 0$, then we obtain\ the canonical semispray associated to
connection $\Gamma $ which is not the same canonical semispray presented by
I. Bucataru and R.~Miron in~[5].

In particular, if $\left( \rho ,\eta \right) =\left( Id_{TM},Id_{M}\right) $%
, $\left( g,h\right) =\left( Id_{E},Id_{M}\right) $, and $F_{e}=0$, then we
obtain the classical canonical semispray associated to connection $\Gamma $.

Using \emph{Theorem 8.1}, we obtain the following:

\textbf{Theorem 8.2 }\emph{The following properties hold good:}\medskip

$1^{\circ }$\ \emph{Since } $\overset{\circ }{\tilde{\delta}}_{c}=\tilde{%
\partial}_{c}-\left( \rho ,\eta \right) \mathring{\Gamma}_{c}^{a}\overset{%
\cdot }{\tilde{\partial}}_{a},~c\in \overline{1,r},$ \emph{it results that }%
\begin{equation*}
\begin{array}[t]{l}
\overset{\circ }{\tilde{\delta}}_{c}=\tilde{\delta}_{c}-\frac{1}{4}\tilde{g}%
_{c}^{b}\circ h\circ \pi \cdot \frac{\partial F^{a}}{\partial y^{b}}\overset{%
\cdot }{\tilde{\partial}}_{a},~c\in \overline{1,r}.%
\end{array}%
\leqno(8.4)
\end{equation*}

$2^{\circ }\ $\emph{Since} $\mathring{\delta}\tilde{y}^{a}=\left( \rho ,\eta
\right) \mathring{\Gamma}_{c}^{a}d\tilde{z}^{c}+d\tilde{y}^{a},$ \emph{it
results that \ \ }%
\begin{equation*}
\begin{array}[t]{l}
\mathring{\delta}\tilde{y}^{a}=\delta \tilde{y}^{a}+\frac{1}{4}\tilde{g}%
_{c}^{b}\circ h\circ \pi \frac{\partial F^{a}}{\partial y^{b}}d\tilde{z}%
^{c},~a\in \overline{1,r}.%
\end{array}%
\leqno(8.5)
\end{equation*}

\textbf{Theorem 8.3 }\emph{The real local functions}%
\begin{equation*}
\begin{array}[t]{l}
\left( \frac{\partial \left( \rho ,\eta \right) \Gamma _{c}^{a}}{\partial
y^{b}},\frac{\partial \left( \rho ,\eta \right) \Gamma _{c}^{a}}{\partial
y^{b}},0,~0\right) ,~a,b,c\in \overline{1,r},%
\end{array}%
\leqno(8.6)
\end{equation*}%
\emph{and}
\begin{equation*}
\begin{array}[t]{l}
\left( \frac{\partial \left( \rho ,\eta \right) \mathring{\Gamma}_{c}^{a}}{%
\partial y^{b}},\frac{\partial \left( \rho ,\eta \right) \mathring{\Gamma}%
_{c}^{a}}{\partial y^{b}},0,~0\right) ,~a,b,c\in \overline{1,r},%
\end{array}%
\leqno(8.6)^{\prime }
\end{equation*}%
\emph{respectively, are the coefficients to a normal Berwald linear }$\left(
\rho ,\eta \right) $\emph{-connection for the generalized tangent bundle }$%
\left( \left( \rho ,\eta \right) TE,\left( \rho ,\eta \right) \tau
_{E},E\right) $.

\textbf{Theorem 8.4 }\emph{The tensor of integrability of the }$\left( \rho
,\eta \right) $\emph{-connection }$\left( \rho ,\eta \right) \mathring{\Gamma%
}$\emph{\ is as follows:}%
\begin{equation*}
\begin{array}{l}
\displaystyle\left( \rho ,\eta ,h\right) \mathbb{\mathring{R}}%
_{cd}^{a}=\left( \rho ,\eta ,h\right) \mathbb{R}_{cd}^{a}+\frac{1}{4}\left(
\tilde{g}_{d}^{e}\circ h\circ \pi \frac{\partial F^{a}}{\partial y^{e}}_{|c}-%
\tilde{g}_{c}^{e}\circ h\circ \pi \frac{\partial F^{a}}{\partial y^{e}}%
_{|d}\right) \vspace*{1mm} \\
\displaystyle+\frac{1}{16}\left( \tilde{g}_{d}^{e}\circ h\circ \pi \frac{%
\partial F^{b}}{\partial y^{e}}\tilde{g}_{c}^{f}\circ h\circ \pi \frac{%
\partial ^{2}F^{a}}{\partial y^{b}\partial y^{f}}-\tilde{g}_{c}^{f}\circ
h\circ \pi \frac{\partial F^{b}}{\partial y^{f}}\tilde{g}_{d}^{e}\circ
h\circ \pi \frac{\partial ^{2}F^{a}}{\partial y^{b}\partial y^{e}}\right)
\vspace*{1mm} \\
\displaystyle+\frac{1}{4}\left( L_{cd}^{f}\circ h\circ \pi \right) \left(
\tilde{g}_{f}^{e}\circ h\circ \pi \right) \frac{\partial F^{a}}{\partial
y^{e}},%
\end{array}%
\leqno(8.7)
\end{equation*}%
\emph{where }$_{|c}$\emph{\ is the }$h$\emph{-covariant derivation with
respect to the normal Berwald linear }$\left( \rho ,\eta \right) $\emph{%
-connection }$(8.6)$\emph{.}

\emph{Proof. }Since
\begin{equation*}
\begin{array}{cl}
\left( \rho ,\eta ,h\right) \mathbb{\mathring{R}}_{cd}^{a}= & \Gamma \left(
\tilde{\rho},Id_{E}\right) \left( \overset{\circ }{\tilde{\delta}}%
_{c}\right) \left( \left( \rho ,\eta \right) \mathring{\Gamma}%
_{d}^{a}\right) -\Gamma \left( \tilde{\rho},Id_{E}\right) \left( \overset{%
\circ }{\tilde{\delta}}_{d}\right) \left( \left( \rho ,\eta \right)
\mathring{\Gamma}_{c}^{a}\right) \\
& +L_{cd}^{e}\circ h\circ \left( h\circ \pi \right) \left( \rho ,\eta
\right) \mathring{\Gamma}_{e}^{a},%
\end{array}%
\end{equation*}%
and%
\begin{equation*}
\begin{array}{cl}
\Gamma \left( \tilde{\rho},Id_{E}\right) \left( \overset{\circ }{\tilde{%
\delta}}_{c}\right) \left( \left( \rho ,\eta \right) \mathring{\Gamma}%
_{d}^{a}\right) & \displaystyle=\Gamma \left( \tilde{\rho},Id_{E}\right)
\left( \tilde{\delta}_{c}\right) \left( \left( \rho ,\eta \right) \Gamma
_{d}^{a}\right) \vspace*{1mm} \\
& \displaystyle+\frac{1}{4}\Gamma \left( \tilde{\rho},Id_{E}\right) \left(
\tilde{\delta}_{c}\right) \left( \tilde{g}_{d}^{e}\circ h\circ \pi \frac{%
\partial F^{a}}{\partial y^{e}}\right) \vspace*{1mm} \\
& \displaystyle-\frac{1}{4}\tilde{g}_{c}^{e}\circ h\circ \pi \frac{\partial
F^{f}}{\partial y^{e}}\frac{\partial }{\partial y^{f}}\left( \left( \rho
,\eta \right) \Gamma _{d}^{a}\right) \vspace*{1mm} \\
& \displaystyle-\frac{1}{16}\tilde{g}_{c}^{e}\circ h\circ \pi \frac{\partial
F^{f}}{\partial y^{e}}\frac{\partial }{\partial y^{f}}\left( \tilde{g}%
_{d}^{e}\circ h\circ \pi \frac{\partial F^{a}}{\partial y^{e}}\right) ,%
\end{array}%
\end{equation*}%
\begin{equation*}
\begin{array}{cl}
\Gamma \left( \tilde{\rho},Id_{E}\right) \left( \overset{\circ }{\tilde{%
\delta}}_{d}\right) \left( \left( \rho ,\eta \right) \mathring{\Gamma}%
_{c}^{a}\right) & \displaystyle=\Gamma \left( \tilde{\rho},Id_{E}\right)
\left( \tilde{\delta}_{d}\right) \left( \left( \rho ,\eta \right) \Gamma
_{c}^{a}\right) \vspace*{1mm} \\
& \displaystyle+\frac{1}{4}\Gamma \left( \tilde{\rho},Id_{E}\right) \left(
\tilde{\delta}_{d}\right) \left( \tilde{g}_{c}^{e}\circ h\circ \pi \frac{%
\partial F^{a}}{\partial y^{e}}\right) \vspace*{1mm} \\
& \displaystyle-\frac{1}{4}\tilde{g}_{d}^{e}\circ h\circ \pi \frac{\partial
F^{f}}{\partial y^{e}}\frac{\partial }{\partial y^{f}}\left( \left( \rho
,\eta \right) \Gamma _{c}^{a}\right) \vspace*{1mm} \\
& \displaystyle-\frac{1}{16}\tilde{g}_{d}^{e}\circ h\circ \pi \frac{\partial
F^{f}}{\partial y^{e}}\frac{\partial }{\partial y^{f}}\left( \tilde{g}%
_{c}^{e}\circ h\circ \pi \frac{\partial F^{a}}{\partial y^{e}}\right) ,%
\end{array}%
\end{equation*}%
\begin{equation*}
\begin{array}{cl}
L_{cd}^{e}\circ h\circ \pi \cdot \left( \rho ,\eta \right) \mathring{\Gamma}%
_{e}^{a} & =L_{cd}^{e}\circ h\circ \pi \cdot \left( \rho ,\eta \right)
\Gamma _{e}^{a}\vspace*{1mm} \\
& \displaystyle+L_{cd}^{e}\circ h\circ \pi \cdot \left( \tilde{g}%
_{e}^{f}\circ h\circ \pi \frac{\partial F^{a}}{\partial y^{f}}\right)%
\end{array}%
\end{equation*}%
it results the conclusion of the theorem.\hfill \emph{q.e.d.}

\textbf{Proposition 8.1 }\emph{If }$S$\emph{\ is the canonical }$\left( \rho
,\eta \right) $\emph{-semispray asso\-cia\-ted to the mechanical }$\left(
\rho ,\eta \right) $\emph{-system }$\left( \left( E,\pi ,M\right)
,F_{e},\left( \rho ,\eta \right) \Gamma \right) $\emph{\ and from }$\mathbf{B%
}^{\mathbf{v}}$\emph{-mor\-phism }$\left( g,h\right) $\emph{,\ then }%
\begin{equation*}
\begin{array}[t]{l}
2G^{a%
{\acute{}}%
}=2G^{a}M_{a}^{a%
{\acute{}}%
}\circ h\circ \pi -\left( g_{b}^{a}\circ h\circ \pi \right) y^{b}\left( \rho
_{a}^{i}\circ h\circ \pi \right) \frac{\partial y^{a%
{\acute{}}%
}}{\partial x^{i}}.%
\end{array}%
\leqno(8.8)
\end{equation*}

\emph{Proof.} Since the Jacobian matrix of coordinates transformation is
\begin{equation*}
\left\Vert
\begin{array}{ll}
\,\ \ \ \ \ \ \ M_{a}^{a%
{\acute{}}%
}\circ h\circ \pi & \,\ 0\vspace*{1mm} \\
\rho _{a}^{i}\circ \left( h\circ \pi \right) \displaystyle\frac{\partial
M_{a}^{a%
{\acute{}}%
}\circ \pi }{\partial x^{i}}y^{a} & M_{a}^{a%
{\acute{}}%
}\circ \pi%
\end{array}%
\right\Vert =\left\Vert
\begin{array}{ll}
\,\ \ \ \ \ \ \ M_{a}^{a%
{\acute{}}%
}\circ h\circ \pi & \,\ 0\vspace*{1mm} \\
\rho _{a}^{i}\circ \left( h\circ \pi \right) \displaystyle\frac{\partial y^{a%
{\acute{}}%
}}{\partial x^{i}} & M_{a}^{a%
{\acute{}}%
}\circ \pi%
\end{array}%
\right\Vert
\end{equation*}%
and
\begin{equation*}
\begin{array}{c}
\left\Vert
\begin{array}{ll}
\,\ \ \ \ \ \ \ M_{a}^{a%
{\acute{}}%
}\circ h\circ \pi & \,\ 0\vspace*{1mm} \\
\rho _{a}^{i}\circ \left( h\circ \pi \right) \displaystyle\frac{\partial y^{a%
{\acute{}}%
}}{\partial x^{i}} & M_{a}^{a%
{\acute{}}%
}\circ \pi%
\end{array}%
\right\Vert \cdot \left(
\begin{array}{l}
\,\ \ \ \left( g_{b}^{a}\circ h\circ \pi \right) y^{b}\vspace*{1mm} \\
-2\left( G^{a}-\displaystyle\frac{1}{4}F^{a}\right)%
\end{array}%
\right) =\left(
\begin{array}{l}
\,\ \ \left( g_{b%
{\acute{}}%
}^{a%
{\acute{}}%
}\circ h\circ \pi \right) y^{b%
{\acute{}}%
}\vspace*{1mm} \\
-2\left( G^{a%
{\acute{}}%
}-\displaystyle\frac{1}{4}F^{a%
{\acute{}}%
}\right)%
\end{array}%
\right) ,%
\end{array}%
\end{equation*}%
the conclusion results immediately.\hfill \emph{q.e.d.}

In the following, we consider a differentiable curve $%
\begin{array}[b]{ccc}
I & ^{\underrightarrow{~c~}} & M%
\end{array}%
$ and its $\left( g,h\right) $-lift $\dot{c}.$

\textbf{Definition 8.3 }If it is verifies the following equality:\textit{\ }%
\begin{equation*}
\begin{array}[t]{l}
\frac{d\dot{c}\left( t\right) }{dt}=\Gamma \left( \tilde{\rho},Id_{E}\right)
S\left( \dot{c}\left( t\right) \right) ,%
\end{array}%
\leqno(8.9)
\end{equation*}%
then we say that \emph{the curve }$\dot{c}$\emph{\ is an integral curve of
the }$\left( \rho ,\eta \right) $\emph{-semispray }$S$\emph{\ of the
mechanical }$\left( \rho ,\eta \right) $\emph{-system }$\left( \left( E,\pi
,M\right) ,F_{e},\left( \rho ,\eta \right) \Gamma \right) $,

\textbf{Theorem 8.5 }\emph{All }$\left( g,h\right) $\emph{-lifts solutions
of the equations:\ }%
\begin{equation*}
\begin{array}[t]{l}
\frac{dy^{a}\left( t\right) }{dt}+2G^{a}\!\circ u\left( c,\dot{c}\right)
\left( x\left( t\right) \right) {=}\frac{1}{2}F^{a}\!\circ u\left( c,\dot{c}%
\right) \left( x\left( t\right) \right) \!,\,a{\in }\overline{1,\!r},%
\end{array}%
\leqno(8.10)
\end{equation*}%
\emph{where }$x\left( t\right) =\left( \eta \circ h\circ c\right) \left(
t\right) ,$ \emph{are } \emph{integral curves of the canonical }$\left( \rho
,\eta \right) $\emph{-semispray asso\-cia\-ted to mechanical }$\left( \rho
,\eta \right) $\emph{-system }$\left( \left( E,\pi ,M\right) ,F_{e},\left(
\rho ,\eta \right) \Gamma \right) $\emph{\ and from locally invertible }$%
\mathbf{B}^{\mathbf{v}}$\emph{-mor\-phism }$\left( g,h\right) .$

\emph{Proof.} Since the equality
\begin{equation*}
\begin{array}[t]{l}
\frac{d\dot{c}\left( t\right) }{dt}=\Gamma \left( \tilde{\rho},Id_{E}\right)
S\left( \dot{c}\left( t\right) \right)%
\end{array}%
\end{equation*}%
is equivalent to
\begin{equation*}
\begin{array}{l}
\displaystyle\frac{d}{dt}((\eta \circ h\circ c)^{i}(t),y^{a}(t))\vspace*{2mm}
\\
\qquad =\displaystyle\left( \rho _{a}^{i}\circ \eta \circ h\circ
c(t)g_{b}^{a}\circ h\circ c(t)y^{b}(t),-2\left( G^{a}-\frac{1}{4}%
F^{a}\right) ((\eta \circ h\circ c)^{i}(t),y^{a}(t))\right) ,%
\end{array}%
\end{equation*}%
it results
\begin{equation*}
\begin{array}{l}
\displaystyle\frac{dy^{a}\left( t\right) }{dt}+2G^{a}\left( x^{i}\left(
t\right) ,y^{a}\left( t\right) \right) =\frac{1}{2}F^{a}\left( x^{i}\left(
t\right) ,y^{a}\left( t\right) \right) ,\ \ a\in \overline{1,n},\vspace*{2mm}
\\
\displaystyle\frac{dx^{i}\left( t\right) }{dt}=\rho _{a}^{i}\circ \eta \circ
h\circ c\left( t\right) g_{b}^{a}\circ h\circ c\left( t\right) y^{b}\left(
t\right) ,%
\end{array}%
\end{equation*}%
where $x^{i}\left( t\right) =\left( \eta \circ h\circ c\right) ^{i}\left(
t\right) $. \hfill \emph{q.e.d.}

\textbf{Definition 8.4 }If $S$\ is a $\left( \rho ,\eta \right) $-semispray,
then the vector field
\begin{equation*}
\begin{array}{l}
\left[ \mathbb{C},S\right] _{\left( \rho ,\eta \right) TE}-S%
\end{array}%
\leqno(8.11)
\end{equation*}%
will be called the \emph{derivation of }$\left( \rho ,\eta \right) $\emph{%
-semispray }$S.$

The $\left( \rho ,\eta \right) $-semispray $S$ will be called $\left( \rho
,\eta \right) $\emph{-spray} if the following conditions are
verified:\medskip

1. $S\circ 0\in C^{1},$\ where $0$\ is the null section;\smallskip

2. Its derivation is the null vector field.\medskip

The $\left( \rho ,\eta \right) $-semispray $S$\ will be called \emph{%
quadratic }$\left( \rho ,\eta \right) $\emph{-spray }if there are verified
the following conditions:\medskip

1. $S\circ 0\in C^{2},$\ where $0$\ is the null section;\smallskip

2. Its derivation is the null vector field.\medskip

In particular, \ if $\ \left( \rho ,\eta \right) =\left(
id_{TM},Id_{M}\right) $ and $\left( g,h\right) =\left( Id_{E},Id_{M}\right)
, $ \ then \ we \ obtain \ the \ \emph{spray} \ and the \emph{quadratic
spray }which is similar with the classical spray and quadratic spray.

\textbf{Theorem 8.6 }\emph{If }$S$\emph{\ is the canonical }$\left( \rho
,\eta \right) $\emph{-spray associated to mechanical }$\left( \rho ,\eta
\right) $\emph{-system }$\left( \left( E,\pi ,M\right) ,F_{e},\left( \rho
,\eta \right) \Gamma \right) $\emph{\ and from locally invertible }$\mathbf{B%
}^{\mathbf{v}}$\emph{-morphism }$\left( g,h\right) $\emph{, then}%
\begin{equation*}
\begin{array}{cl}
2\left( G^{a}-\frac{1}{4}F^{a}\right)  & =\left( \rho ,\eta \right) \Gamma
_{c}^{a}\left( g_{f}^{c}\circ h\circ \pi \right) y^{f} \\
& +\frac{1}{2}\left( g_{e}^{d}\circ h\circ \pi \right) y^{e}\left(
L_{dc}^{b}\circ h\circ \pi \right) \left( \tilde{g}_{b}^{a}\circ h\circ \pi
\right) \left( g_{f}^{c}\circ h\circ \pi \right) y^{f} \\
& -\frac{1}{2}\left( \rho _{c}^{j}\circ h\circ \pi \right) \frac{\partial
\left( g_{e}^{b}\circ h\circ \pi \right) }{\partial x^{j}}y^{e}\left( \tilde{%
g}_{b}^{a}\circ h\circ \pi \right) \left( g_{f}^{c}\circ h\circ \pi \right)
y^{f} \\
& +\frac{1}{2}\left( g_{e}^{b}\circ h\circ \pi \right) y^{e}\left( \rho
_{b}^{i}\circ h\circ \pi \right) \frac{\partial \left( \tilde{g}%
_{c}^{a}\circ h\circ \pi \right) }{\partial x^{i}}\left( g_{f}^{c}\circ
h\circ \pi \right) y^{f}%
\end{array}%
\leqno(8.12)
\end{equation*}

\emph{We obtain the spray}%
\begin{equation*}
\begin{array}{cl}
S & =\left( g_{b}^{a}\circ h\circ \pi \right) y^{b}\tilde{\partial}%
_{a}-\left( \rho ,\eta \right) \Gamma _{c}^{a}\left( g_{f}^{c}\circ h\circ
\pi \right) y^{f}\overset{\cdot }{\tilde{\partial}}_{a} \\
& -\frac{1}{2}\left( g_{e}^{d}\circ h\circ \pi \right) y^{e}\left(
L_{dc}^{b}\circ h\circ \pi \right) \left( \tilde{g}_{b}^{a}\circ h\circ \pi
\right) \left( g_{f}^{c}\circ h\circ \pi \right) y^{f}\overset{\cdot }{%
\tilde{\partial}}_{a} \\
& +\frac{1}{2}\left( \rho _{c}^{j}\circ h\circ \pi \right) \frac{\partial
\left( g_{e}^{b}\circ h\circ \pi \right) }{\partial x^{j}}y^{e}\left( \tilde{%
g}_{b}^{a}\circ h\circ \pi \right) \left( g_{f}^{c}\circ h\circ \pi \right)
y^{f}\overset{\cdot }{\tilde{\partial}}_{a} \\
& -\frac{1}{2}\left( g_{e}^{b}\circ h\circ \pi \right) y^{e}\left( \rho
_{b}^{i}\circ h\circ \pi \right) \frac{\partial \left( \tilde{g}%
_{c}^{a}\circ h\circ \pi \right) }{\partial x^{i}}\left( g_{f}^{c}\circ
h\circ \pi \right) y^{f}\overset{\cdot }{\tilde{\partial}}_{a}%
\end{array}%
\leqno(8.13)
\end{equation*}

\emph{This spray will be called the canonical }$\left( \rho ,\eta \right) $%
\emph{-spray associated to mechanical system }$\left( \left( E,\pi ,M\right)
,F_{e},\left( \rho ,\eta \right) \Gamma \right) $\emph{\ and from locally
invertible }$\mathbf{B}^{\mathbf{v}}$\emph{-morphism }$(g,h).$

\emph{In particular, if }$\left( \rho ,\eta \right) =\left(
id_{TM},Id_{M}\right) $\emph{\ and }$\left( g,h\right) =\left(
Id_{E},Id_{M}\right) ,$\emph{\ then we get the canonical spray associated to
connection }$\Gamma $\emph{\ which is similar with the classical canonical
spray associated to connection }$\Gamma $.

\emph{Proof.} Since
\begin{equation*}
\begin{array}[t]{l}
\left[ \mathbb{C},S\right] _{\left( \rho ,\eta \right) TE}=\left[ y^{a}%
\overset{\cdot }{\tilde{\partial}}_{a},\left( g_{e}^{b}\circ h\circ \pi
\cdot y^{e}\right) \tilde{\partial}_{b}\right] _{\left( \rho ,\eta \right)
TE}-2\left[ y^{a}\overset{\cdot }{\tilde{\partial}}_{a},\left( G^{b}-\frac{1%
}{4}F^{b}\right) \overset{\cdot }{\tilde{\partial}}_{b}\right] _{\left( \rho
,\eta \right) TE},%
\end{array}%
\end{equation*}

\begin{equation*}
\!\!%
\begin{array}{cl}
\left[ y^{a}\overset{\cdot }{\tilde{\partial}}_{a},\left( g_{e}^{b}\circ
h\circ \pi \cdot y^{e}\right) \tilde{\partial}_{b}\right] _{\left( \rho
,\eta \right) TE}\!\!\!\! & \displaystyle=\left( g_{e}^{b}\circ h\circ \pi
\cdot y^{e}\right) \tilde{\partial}_{b}\vspace*{2mm}%
\end{array}%
\end{equation*}%
and
\begin{equation*}
\begin{array}{cl}
\left[ y^{a}\overset{\cdot }{\tilde{\partial}}_{a},\left( G^{b}-\frac{1}{4}%
F^{b}\right) \overset{\cdot }{\tilde{\partial}}_{b}\right] _{\left( \rho
,\eta \right) TE} & \displaystyle=y^{a}\frac{\partial \left( G^{b}-\frac{1}{4%
}F^{b}\right) }{\partial y^{a}}\overset{\cdot }{\tilde{\partial}}_{b}-\left(
G^{b}-\frac{1}{4}F^{b}\right) \overset{\cdot }{\tilde{\partial}}_{b}\vspace*{%
2mm}%
\end{array}%
\end{equation*}%
it results that
\begin{equation*}
\begin{array}{cc}
\left[ \mathbb{C},S\right] _{\left( \rho ,\eta \right) TE}-S & \displaystyle%
=2\left( -y^{f}\frac{\partial \left( G^{a}-\frac{1}{4}F^{a}\right) }{y^{f}}%
+2\left( G^{a}-\frac{1}{4}F^{a}\right) \right) \overset{\cdot }{\tilde{%
\partial}}_{a}%
\end{array}%
\leqno\left( S_{1}\right)
\end{equation*}

Using equality $(8.3)$, it results that%
\begin{equation*}
\begin{array}{cl}
\displaystyle\frac{\partial \left( G^{a}-\frac{1}{4}F^{a}\right) }{y^{f}} &
=\left( \rho ,\eta \right) \Gamma _{c}^{a}\left( g_{f}^{c}\circ h\circ \pi
\right) \\
& +\frac{1}{2}\left( g_{e}^{d}\circ h\circ \pi \right) y^{e}\left(
L_{dc}^{b}\circ h\circ \pi \right) \left( \tilde{g}_{b}^{a}\circ h\circ \pi
\right) \left( g_{f}^{c}\circ h\circ \pi \right) \\
& -\frac{1}{2}\left( \rho _{c}^{j}\circ h\circ \pi \right) \frac{\partial
\left( g_{e}^{b}\circ h\circ \pi \right) }{\partial x^{j}}y^{e}\left( \tilde{%
g}_{b}^{a}\circ h\circ \pi \right) \left( g_{f}^{c}\circ h\circ \pi \right)
\\
& +\frac{1}{2}\left( g_{e}^{b}\circ h\circ \pi \right) y^{e}\left( \rho
_{b}^{i}\circ h\circ \pi \right) \frac{\partial \left( \tilde{g}%
_{c}^{a}\circ h\circ \pi \right) }{\partial x^{i}}\left( g_{f}^{c}\circ
h\circ \pi \right)%
\end{array}%
\leqno\left( S_{2}\right)
\end{equation*}

Using equalities $\left( S_{1}\right) $ and $\left( S_{2}\right) $, it
results the conclusion of the theorem.\hfill \emph{q.e.d.}

\textbf{Theorem 8.7 }\emph{\ All }$\left( g,h\right) $\emph{-lifts solutions
of the following system of equations:\ }%
\begin{equation*}
\begin{array}{l}
\displaystyle\frac{dy^{a}}{dt}+\left( \rho ,\eta \right) \Gamma
_{c}^{a}\left( g_{f}^{c}\circ h\circ \pi \right) y^{f}\vspace*{2mm} \\
\displaystyle+\frac{1}{2}\left( g_{e}^{d}\circ h\circ \pi \right)
y^{e}\left( L_{dc}^{b}\circ h\circ \pi \right) \left( \tilde{g}_{b}^{a}\circ
h\circ \pi \right) \left( g_{f}^{c}\circ h\circ \pi \right) y^{f} \\
\displaystyle-\frac{1}{2}\left( \rho _{c}^{j}\circ h\circ \pi \right) \frac{%
\partial \left( g_{e}^{b}\circ h\circ \pi \right) }{\partial x^{j}}%
y^{e}\left( \tilde{g}_{b}^{a}\circ h\circ \pi \right) \left( g_{f}^{c}\circ
h\circ \pi \right) y^{f} \\
\displaystyle+\frac{1}{2}\left( g_{e}^{b}\circ h\circ \pi \right)
y^{e}\left( \rho _{b}^{i}\circ h\circ \pi \right) \frac{\partial \left(
\tilde{g}_{c}^{a}\circ h\circ \pi \right) }{\partial x^{i}}\left(
g_{f}^{c}\circ h\circ \pi \right) y^{f}=0,%
\end{array}%
\leqno(8.14)
\end{equation*}%
\emph{are the integral curves of canonical }$\left( \rho ,\eta \right) $%
\emph{-spray associated to mechanical }$\left( \rho ,\eta \right) $\emph{%
-system }$\left( \left( E,\pi ,M\right) ,F_{e},\left( \rho ,\eta \right)
\Gamma \right) $\emph{\ and from locally invertible }$\mathbf{B}^{\mathbf{v}}
$\emph{-morphism\ }$\left( g,h\right) .$

\section{A Lagrangian formalism for Lagrange mechanical $\left( \protect\rho %
,\protect\eta \right) $-systems}

Let $\left( \left( E,\pi ,M\right) ,F_{e},L\right) $ be an arbitrarily
Lagrange mechanical $\left( \rho ,\eta \right) $-system.

Let $\left( d\tilde{z}^{\alpha },d\tilde{y}^{a}\right) $ be the natural dual
$\left( \rho ,\eta \right) $\emph{-base} of the natural $\left( \rho ,\eta
\right) $-base $\left( \tilde{\partial}_{\alpha },\overset{\cdot }{\tilde{%
\partial}}_{a}\right) .$

It is very important to remark that the $1$-forms $d\tilde{z}^{a},d\tilde{y}%
^{a},~a\in \overline{1,p}$ are not the differentials of coordinates
functions as in the classical case, but we will use the same notations. In
this case
\begin{equation*}
\left( d\tilde{z}^{a}\right) \neq d^{\left( \rho ,\eta \right) TE}\left(
\tilde{z}^{a}\right) ,
\end{equation*}%
where $d^{\left( \rho ,\eta \right) TE}$ is the exterior differentiation
operator associated to exterior differential $\mathcal{F}\left( E\right) $%
-algebra
\begin{equation*}
\left( \Lambda \left( \left( \rho ,\eta \right) TE,\left( \rho ,\eta \right)
\tau _{E},E\right) ,+,\cdot ,\wedge \right) .
\end{equation*}

Let $L$ be a regular Lagrangian and let $\left( g,h\right) $\ be a locally
invertible $\mathbf{B}^{\mathbf{v}}$-morphism of $\left( E,\pi ,M\right) $
source and target.

\textbf{Definition 9.1 } The $1$-form
\begin{equation*}
\begin{array}{c}
\theta _{L}=\left( \tilde{g}_{a}^{e}\circ h\circ \pi \cdot L_{e}\right) d%
\tilde{z}^{a}%
\end{array}%
\leqno(9.1)
\end{equation*}%
will be called the $1$\emph{-form of Poincar\'{e}-Cartan type associated to
the Lagrangian }$L$ \emph{and to the locally invertible }$\mathbf{B}^{%
\mathbf{v}}$\emph{-morphism }$\left( g,h\right) $.\medskip

We obtain easily:
\begin{equation*}
\begin{array}[t]{l}
\theta _{L}\left( \tilde{\partial}_{a}\right) =\tilde{g}_{b}^{e}\circ h\circ
\pi \cdot L_{e},\,\,\ \theta _{L}\left( \overset{\cdot }{\tilde{\partial}}%
_{b}\right) =0.%
\end{array}%
\leqno(9.2)
\end{equation*}

\textbf{Definition 9.2 } The $2$-form
\begin{equation*}
\omega _{L}=d^{\left( \rho ,\eta \right) TE}\theta _{L}
\end{equation*}%
will be called the $2$\emph{-form of Poincar\'{e}-Cartan type associated to
the Lagrangian }$L$\emph{\ and to the locally invertible }$\mathbf{B}^{%
\mathbf{v}}$\emph{-morphism }$\left( g,h\right) $.\medskip

By the definition of $d^{\left( \rho ,\eta \right) TE},$ we obtain:
\begin{equation*}
\begin{array}{ll}
\omega _{L}\left( U,V\right) & \displaystyle=\Gamma \left( \tilde{\rho}%
,Id_{E}\right) \left( U\right) \left( \theta _{L}\left( V\right) \right)
\vspace*{2mm} \\
& \displaystyle-\,\Gamma \left( \tilde{\rho},Id_{E}\right) \left( V\right)
\left( \theta _{L}\left( U\right) \right) -\theta _{L}\left( \left[ U,V%
\right] _{\left( \rho ,\eta \right) TE}\right) ,%
\end{array}%
\leqno(9.3)
\end{equation*}%
for any $U,V\in \Gamma \left( \left( \rho ,\eta \right) TE,\left( \rho ,\eta
\right) \tau _{E},E\right) $.\smallskip

\textbf{Definition 9.3 } The real function
\begin{equation*}
\begin{array}{c}
\mathcal{E}_{L}=y^{a}L_{a}-L%
\end{array}%
\leqno(9.4)
\end{equation*}%
will be called the \emph{energy of regular Lagrangian }$L.$

\textbf{Theorem 9.1 }\emph{The equation }%
\begin{equation*}
\begin{array}{c}
i_{S}\left( \omega _{L}\right) =-d^{\left( \rho ,\eta \right) TE}\left(
\mathcal{E}_{L}\right) ,\,S\in \Gamma \left( \left( \rho ,\eta \right)
TE,\left( \rho ,\eta \right) \tau _{E},E\right) ,%
\end{array}%
\leqno\left( 9.5\right)
\end{equation*}%
\emph{has an unique solution }$S_{L}\left( g,h\right) $\emph{\ of the type: }%
\begin{equation*}
\begin{array}[t]{l}
\left( g_{e}^{a}\circ h\circ \pi \right) y^{e}\tilde{\partial}_{a}-2\left(
G^{a}-\frac{1}{4}F^{a}\right) \overset{\cdot }{\tilde{\partial}}_{a},%
\end{array}%
\leqno(9.6)
\end{equation*}%
\emph{where }%
\begin{equation*}
\begin{array}[t]{l}
-2\left( G^{a}-\frac{1}{4}F^{a}\right) =E_{b}\left( L,g,h\right) \tilde{L}%
^{be}\left( g_{e}^{a}\circ h\circ \pi \right)
\end{array}%
\leqno(9.7)
\end{equation*}%
\emph{and}%
\begin{equation*}
\begin{array}{cl}
E_{b}\left( L,g,h\right)  & =\left( \rho _{b}^{i}{\circ }h{\circ }\pi
\right) L_{i}-\left( \rho _{b}^{i}{\circ }h{\circ }\pi \right) y^{a}L_{ia}
\\
& -\left( g_{f}^{d}\circ h\circ \pi \right) y^{f}\left( \rho _{d}^{i}{\circ }%
h{\circ }\pi \right) \frac{\partial \left( \left( \tilde{g}_{b}^{e}\circ
h\circ \pi \right) L_{e}\right) }{\partial x^{i}} \\
& +\left( g_{f}^{d}\circ h\circ \pi \right) y^{f}\left( \rho _{b}^{i}{\circ }%
h{\circ }\pi \right) \frac{\partial \left( \left( \tilde{g}_{d}^{e}\circ
h\circ \pi \right) L_{e}\right) }{\partial x^{i}} \\
& +\left( g_{f}^{d}\circ h\circ \pi \right) y^{f}\left( L_{db}^{c}{\circ }h{%
\circ }\pi \right) \left( \tilde{g}_{c}^{e}\circ h\circ \pi \right) L_{e}%
\end{array}%
\hspace*{-4mm}\leqno(9.8)
\end{equation*}

$S_{L}\left( g,h\right) $\textit{\ }will be called \emph{the\ canonical }$%
\left( \rho ,\eta \right) $\emph{-semispray associated to Lagrange
mechanical }$\left( \rho ,\eta \right) $\emph{-system }$\left( \left( E,\pi
,M\right) ,F_{e},L\right) $\emph{\ and from locally invertible }$\mathbf{B}^{%
\mathbf{v}}$\emph{-morphism }$(g,h).$

\emph{Proof.} We obtain that
\begin{equation*}
i_{S}\left( \omega _{L}\right) =-d^{\left( \rho ,\eta \right) TE}\left(
\mathcal{E}_{L}\right)
\end{equation*}%
if and only if
\begin{equation*}
\omega _{L}\left( S,X\right) =-\Gamma \left( \tilde{\rho},Id_{E}\right)
\left( X\right) \left( \mathcal{E}_{L}\right) ,
\end{equation*}%
for any $X\in \Gamma \left( \left( \rho ,\eta \right) TE,\left( \rho ,\eta
\right) \tau _{E},E\right) .$

Particularly, we obtain:%
\begin{equation*}
\begin{array}[t]{l}
\omega _{L}\left( S,\tilde{\partial}_{b}\right) =-\Gamma \left( \tilde{\rho}%
,Id_{E}\right) \left( \tilde{\partial}_{b}\right) \left( \mathcal{E}%
_{L}\right) .%
\end{array}%
\end{equation*}

If we expand this equality, we obtain%
\begin{equation*}
\begin{array}{l}
\left( g_{f}^{d}\circ h\circ \pi \right) y^{f}\left[ \left( \rho _{d}^{i}{%
\circ }h{\circ }\pi \right) \frac{\partial \left( \left( \tilde{g}%
_{b}^{e}\circ h\circ \pi \right) L_{e}\right) }{\partial x^{i}}-\left( \rho
_{b}^{i}{\circ }h{\circ }\pi \right) \frac{\partial \left( \left( \tilde{g}%
_{d}^{e}\circ h\circ \pi \right) L_{e}\right) }{\partial x^{i}}\right. \\
\displaystyle\left. -\left( L_{ab}^{c}{\circ }h{\circ }\pi \right) \left(
\tilde{g}_{c}^{e}\circ h\circ \pi \right) L_{e}\right] -2\left( G^{a}-\frac{1%
}{4}F^{a}\right) \left( \tilde{g}_{a}^{e}\circ h\circ \pi \right) \cdot
L_{eb}\vspace*{2mm} \\
\qquad \displaystyle=\rho _{b}^{i}{\circ }h{\circ }\pi \cdot L_{i}-\left(
\rho _{b}^{i}{\circ }h{\circ }\pi \right) \frac{\partial \left(
y^{a}L_{a}\right) }{\partial x^{i}}.%
\end{array}%
\end{equation*}

After some calculations, we obtain the conclusion of the theorem.\hfill
\emph{q.e.d.}

\emph{Remark 9.1 }If $F_{e}=0$ and $\eta =Id_{M},$\ then
\begin{equation*}
\begin{array}{cl}
E_{b}\left( L,Id_{E},Id_{M}\right) & =\left( \rho _{b}^{i}{\circ }\pi
\right) L_{i}-\left( \rho _{b}^{i}{\circ }\pi \right)
y^{d}L_{id}+y^{d}\left( L_{db}^{c}{\circ }\pi \right) L_{c}%
\end{array}%
\hspace*{-4mm}
\end{equation*}%
and $S_{L}\left( Id_{E},Id_{M}\right) \overset{put}{=}S_{L}$\ is the
canonical $\rho $-semispray associated to regular Lagrangian $L$\ which is
similar with the semispray presented in $\left[ 9\right] $ by M. de Leon, J.
Marrero and E. Martinez. (see also $\left[ 11\right] $)

In addition, if $F_{e}\neq 0$ and\ $\left( \rho ,\eta \right) =\left(
Id_{TM},Id_{M}\right) $, then $S_{L}\left( Id_{E},Id_{M}\right) \overset{put}%
{=}S_{L}$\ will be called \emph{the canonical semispray}\ which is similar
with the semispray presented by I.~Bucataru and R. Miron in $\left[ 5\right]
$.

In particular, if $F_{e}=0$\ and $\left( \rho ,\eta \right) =\left(
Id_{TM},Id_{M}\right) $, then $S_{L}\left( Id_{M},Id_{E}\right) \overset{put}%
{=}S_{L}$\ will be called \emph{the canonical semispray}\ which is similar
with the canonical semispray presented by R.~Miron and M. Anastasiei in~$%
[13].$ (see also $\left[ 14,15\right] $)

\textbf{Theorem 9.2}\textit{\ }\emph{If }$S_{L}\left( g,h\right) $\textit{\ }%
\emph{is the\ canonical }$\left( \rho ,\eta \right) $\emph{-semispray
associated to Lagrange mechanical }$\left( \rho ,\eta \right) $\emph{-system
}$\left( \left( E,\pi ,M\right) ,F_{e},L\right) $\emph{\ and from locally
invertible }$\mathbf{B}^{\mathbf{v}}$\emph{-morphism }$(g,h),$ \emph{then
the\ real local functions }%
\begin{equation*}
\begin{array}{cl}
\left( \rho ,\eta \right) \Gamma _{c}^{a} & =-\frac{1}{2}\left( \tilde{g}%
_{c}^{d}\circ h\circ \pi \right) \frac{\partial \left( E_{b}\left(
L,g,h\right) \tilde{L}^{be}\left( g_{e}^{a}\circ h\circ \pi \right) \right)
}{\partial y^{d}} \\
& -\frac{1}{2}\left( g_{e}^{d}\circ h\circ \pi \right) y^{e}\left(
L_{dc}^{f}\circ h\circ \pi \right) \left( \tilde{g}_{f}^{a}\circ h\circ \pi
\right) \\
& +\frac{1}{2}\left( \rho _{c}^{j}\circ h\circ \pi \right) \frac{\partial
\left( g_{e}^{b}\circ h\circ \pi \right) }{\partial x^{j}}y^{e}\left( \tilde{%
g}_{b}^{a}\circ h\circ \pi \right) \\
& -\frac{1}{2}\left( g_{e}^{b}\circ h\circ \pi \right) y^{e}\left( \rho
_{b}^{i}\circ h\circ \pi \right) \frac{\partial \left( \tilde{g}%
_{c}^{a}\circ h\circ \pi \right) }{\partial x^{i}}%
\end{array}%
\leqno(9.9)
\end{equation*}%
\emph{are the components of a }$\left( \rho ,\eta \right) $\emph{-connection
}$\left( \rho ,\eta \right) \Gamma $\emph{\ for the vector bundle }$\left(
E,\pi ,M\right) $\emph{\ which will be called the }$\left( \rho ,\eta
\right) $\emph{-connection associated to Lagrange mechanical }$\left( \rho
,\eta \right) $\emph{-system}\break $\left( \left( E,\pi ,M\right)
,F_{e},L\right) $\emph{\ and from locally invertible }$\mathbf{B}^{\mathbf{v}%
}$\emph{-morphism} $(g,h).$

\emph{In particular, if }$\eta =h=Id_{M}$\emph{\ and }$g=Id_{E},$\emph{\
then we obtain}
\begin{equation*}
\begin{array}{cl}
\rho \Gamma _{c}^{a} & =\displaystyle-\frac{1}{2}\frac{\partial \left(
E_{b}\left( L,Id_{E},Id_{M}\right) \tilde{L}^{ba}\right) }{\partial y^{c}}-%
\frac{1}{2}y^{b}L_{bc}^{a}\circ \pi .%
\end{array}%
\leqno(9.9^{\prime })
\end{equation*}

\textbf{Theorem 9.3 }\emph{The parallel }$\left( g,h\right) $\emph{-lifts
with respect to }$\left( \rho ,\eta \right) $\emph{-connection }$\left( \rho
,\eta \right) \Gamma $ \emph{are the\ integral curves of the canonical }$%
\left( \rho ,\eta \right) $\emph{-semispray associated to\ mechanical }$%
\left( \rho ,\eta \right) $\emph{-system }$\left( \left( E,\pi ,M\right)
,F_{e},L\right) $ \emph{and from locally invertible }$\mathbf{B}^{\mathbf{v}%
} $\emph{-morphism }$\left( g,h\right) .$

\textbf{Definition 9.4 }The equations
\begin{equation*}
\begin{array}{c}
\,\dfrac{dy^{a}\left( t\right) }{dt}-\left( E_{b}\left( L,g,h\right) \tilde{L%
}^{be}\left( g_{e}^{a}\circ h\circ \pi \right) \right) \circ u\left( c,\dot{c%
}\right) \left( x\left( t\right) \right) =0,%
\end{array}%
\leqno(9.10)
\end{equation*}%
where $x\left( t\right) =\eta \circ h\circ c\left( t\right) $, will be
called the \emph{equations of Euler-Lagrange type associated to Lagrange
mechanical }$\left( \rho ,\eta \right) $\emph{-system }$\left( \left( E,\pi
,M\right) ,F_{e},L\right) $\emph{\ and from locally invertible }$\mathbf{B}^{%
\mathbf{v}}$\emph{-morphism }$\left( g,h\right) .$

The equations
\begin{equation*}
\begin{array}{c}
\dfrac{dy^{a}\left( t\right) }{dt}-\left( E_{b}\left( L,Id_{E},Id_{M}\right)
\tilde{L}^{ba}\right) \circ u\left( c,\dot{c}\right) \left( x\left( t\right)
\right) =0,%
\end{array}%
\leqno(9.10^{\prime })
\end{equation*}%
where $x\left( t\right) =c\left( t\right) $, will be called the \emph{%
equations of Euler-Lagrange type associated to Lagrange mechanical} $\left(
\rho ,\eta \right) $\emph{-system }$\left( \left( E,\pi ,M\right)
,F_{e},L\right) $.

\emph{Remark 9.1 }The\ integral curves of the canonical $\left( \rho ,\eta
\right) $-semispray associated to mechanical $\left( \rho ,\eta \right) $%
-system $\left( \left( E,\pi ,M\right) ,F_{e},L\right) $\ and from locally
invertible\emph{\ }$\mathbf{B}^{\mathbf{v}}$-morphism $\left( g,h\right) $\
are the $\left( g,h\right) $-lifts solutions for the equations of
Euler-Lagrange type $\left( 9.10\right) $.

Using our theory, we obtain the following

\textbf{Theorem 9.4 }\emph{If }$F$\emph{\ is a Finsler fundamental function,
then the\ geodesics on the manifold }$M$\emph{\ are the curves such that the
components of their }$\left( g,h\right) $\emph{-lifts are\ solutions for the
equations of Euler-Lagrange type }$\left( 9.10\right) .$\bigskip

Therefore, it is natural to propose to extend the study of the Finsler
geometry from the usual Lie algebroid $\left( \left( TM,\tau _{M},M\right) ,%
\left[ ,\right] _{TM},\left( Id_{TM},Id_{M}\right) \right) ,$ to an
arbitrary (generalized) Lie algebroid $\left( \left( E,\pi ,M\right) ,\left[
,\right] _{E,h},\left( \rho ,\eta \right) \right) .$

\bigskip

\hfill
\begin{tabular}{c}
SECONDARY SCHOOL \textquotedblleft CORNELIUS RADU\textquotedblright , \\
RADINESTI VILLAGE, 217196, GORJ COUNTY, ROMANIA \\
e-mail: c\_arcus@yahoo.com, c\_arcus@radinesti.ro%
\end{tabular}


\addcontentsline{toc}{section}{References}

\begin{thebibliography}{99}
\bibitem{e1} C. M. Arcu\c{s}, \emph{Algebraic constructions in the category
of Lie algebroids}, arXiv:math.DG/1101.0960v3, 20 Jul (2011).

\bibitem{e2} C. M. Arcu\c{s}, \emph{Algebraic constructions in the category
of vector bundles, }arXiv: math. DG/1101.0956 v3, 25 Jul (2011).

\bibitem{e3} C. M. Arcu\c{s}, \emph{Interior and exterior differential
systems for Lie algebroids}, Advances in Pure Mathematics (accepted).

\bibitem{e4} P.L. Antonelli, R.S. Ingarden, M. Matsumoto, \emph{The Theory
of Sprays and Finsler Spaces with Applications in Physics and Biology,}
Fundamental Theories of Physics, Kluwer Academic Publisher, FTPH no.\textbf{%
\ 58,} Dordrecht, (1993).

\bibitem{e5} I. Buc\u{a}taru and R. Miron, \emph{Finsler-Lagrange Geometry.
Applications to dynamical systems,} Ed Academiei Romane, Bucuresti, (2007).

\bibitem{e6} J. Grabowski, P. Urbanski, \emph{Lie algebroids and
Poisson-Nijenhuis structures,} Rep. Math. Phys. \textbf{40}, (1997), 196-208.

\bibitem{e7} J. Kern, \emph{Lagrange geometry}, Arch. Math., \textbf{25},
(1974), 438-443.

\bibitem{e8} J. Klein, \emph{Espaces variationales et mecanique,} Ann. Inst.
Fourier, \textbf{12}, (1962), 1-124.

\bibitem{e9} M. de Leon, J. Marrero, E. Martinez, \emph{Lagrangian
submanifolds and dynamics on Lie algebroids,} arXiv: math. DG/0407528 v1,
(2004).

\bibitem{e10} P. Libermann, \emph{Lie algebroids and Mechanics,} Arch. Math.
(Brno), \textbf{32}, (1996), 147-162.

\bibitem{e11} E. Martinez, \emph{Lagrangian Mechanics on Lie algebroids},
Acta Aplicadae Ma\-te\-ma\-ticae, \textbf{67}, (2001), 295-320.

\bibitem{e12} R. Miron, \emph{A Lagrangian theory of Relativity}, (I, II)
Ann. St. Univ. Al. I. Cuza, Iasi, XXXII, s.1., f.2, f.3, \textbf{32},
(1986), 37-62, 7-16.

\bibitem{e13} R, Miron and M. Anastasiei, \emph{The Geometry of Lagrange
spaces. Theory and applications}, Kluwer Academic Publishers, FTPH no.
\textbf{59}, (1994).

\bibitem{e14} R. Miron, M. Anastasiei, I. Buc\u{a}taru, \emph{The Geometry
of Lagrange Spaces,} In Antonelli P.L. (ed.), Handbook of Finsler Geometry,
Kluwer Acad. Publ., (2003), 969-1124.

\bibitem{e15} R, Miron, Drago\c{s} Hrimiuc, Hideo Shimada, Sorin V. Sabau,
\emph{The Geometry of Hamilton and Lagrange Spaces}, Kluver Academic
Publishers, FTPH 118, 2001.

\bibitem{e16} L. Popescu, \emph{The geometry of Lie algebroids and
applications to optimal control,} Annals. Univ. Al. I. Cuza, Iasi, Series I,
Math., LI, 95-109, (2005).

\bibitem{e17} L. Popescu, \emph{Geometrical structures on Lie algebroids,}
Publicationes Mathematicae Debreten, \textbf{72}, 1-2, 95-109, (2008).

\bibitem{e18} J. Szilasi, \emph{The Geometry of Lagrange Spaces,} In
Antonelli P.L. (ed.), Handbook of Finsler Geometry, Kluwer Acad. Publ.,
(2003), 1185-1426.

\bibitem{e19} S. Vacaru, \emph{Clifford Algebroids and Nonholonomic Spinor
Deformations of Taub-NUT Spacetimes,} ArXiv: hep-th/0502145v1, (2005).

\bibitem{e20} S. Vacaru, \emph{Nonholonomic Deformations of Disk Solutions
and Algebroid Symmetries in Einstein and Extra Dimension Gravity,} ArXiv:
gr-qc/0504095v1, (2005).

\bibitem{e21} S. Vacaru, \emph{Clifford-Finsler algebroids and nonholonomic
Einstein-Dirac structures,} J. of Math. Phys. \textbf{47}, 2093504,1-20,
(2006).

\bibitem{e22} S. Vacaru, \emph{Nonholonomic Algebroids, Finsler Geometry and
Lagrange-Hamilton Spaces,} ArXiv: math-ph/0705.0032v1, (2007).

\bibitem{e23} A. Weinstein, \emph{Lagrangian mechanics and groupoids}, Field
Institute Communications, \textbf{7}, (1996), 206-231.
\end{thebibliography}
\end{document}